%% file: paper.tex
\documentclass[a4paper]{aa}

\usepackage{natbib} 
\usepackage[english]{babel}
\usepackage{latexsym}
\usepackage{graphicx}
\usepackage{epsf} 
\usepackage[intlimits,sumlimits]{amsmath}

\input{Befehle}

\renewcommand{\Abl}[2]{\frac{{\rm d} #1}{{\rm d} #2}}
\newcommand{\thr}{\theta_{\rm r}}
\newcommand{\yd}{y_{\rm d}}
\newcommand{\yo}{y_{\rm opt}}
\newcommand{\yc}{y_{\rm r}}
\newcommand{\yb}{y_{\rm b}}
\newcommand{\ybr}{y_{\rm b,r}}
\newcommand{\yr}{y_{\rm r}}
\newcommand{\yref}{{y_{\rm ref}}}
\newcommand{\ysum}{y_{\rm sum}}
\newcommand{\ydif}{y_{\rm dif}}
\newcommand{\ydifb}{y_{\rm b}}
\newcommand{\ydifs}{y_{\rm sup, 12}}
\newcommand{\ydifrs}{y_{\rm r, 12}}
\newcommand{\yxxx}{y_{12}}
\newcommand{\Deltas}{\Delta_{\rm sup, 12}}
\newcommand{\Deltars}{\Delta_{\rm r, 12}}
\newcommand{\DeltaRJ}{\Delta_{\rm RJ}}

\newcommand{\Deltar}{\Delta_{\rm r}}

\newcommand{\bd}{\beta_{\rm d}}

\newcommand{\Ir}{I_{\rm ref}}

\newcommand{\id}{{\,\rm d}}
\newcommand{\TRJ}{T_{\mathrm{RJ}}}
\newcommand{\Tbar}{\bar{T}}
\newcommand{\Tref}{T_{\rm ref}}
\newcommand{\Treft}{T'_{\rm ref}}
\newcommand{\Tast}{T_\ast}

\newcommand{\WMAP}{{\sc Wmap }}
\newcommand{\defSB}{{SB}}

\newcommand{\SoBs}{{SB }}

\newcommand{\lesssim}{\mathrel{\hbox{\rlap{\hbox{\lower4pt\hbox{$\sim$}}}\hbox{$<$}}}}
\newcommand{\gtrsim}{\mathrel{\hbox{\rlap{\hbox{\lower4pt\hbox{$\sim$}}}\hbox{$>$}}}} 
\newcommand{\plotwd}{8.6cm}
\newcommand{\Ipl}{B_{\nu}}
\newcommand{\Is}{I_{\rm S}}

\newcommand{\Iref}{I_{\rm ref}}
\newcommand{\x}{{\hat{x}}}
\newcommand{\xn}{x}
\renewcommand{\xc}{\hat{x}_{\rm c}}
\renewcommand{\kB}{k\,}

\begin{document}

%\title{Unavoidable spectral distortions of the cosmic microwave background due
%to existence of angular temperature fluctuations}
%\titlerunning{Unavoidable spectral distortions of the cosmic microwave background} 
%\title{CMB spectral distortions due to the observed angular
%  temperature fluctuations}
\title{Superposition of blackbodies and the dipole anisotropy: \\A
possibility to calibrate CMB experiments}

\titlerunning{Superposition of blackbodies and the dipole anisotropy}

\author{J. Chluba\inst{1} \and R.A. Sunyaev\inst{1,2}}

\institute{Max-Planck-Institut f\"ur Astrophysik, Karl-Schwarzschild-Str. 1,
86740 Garching bei M\"unchen, Germany \and 
Space Research Institute, Russian Academy of Sciences, Profsoyuznaya 84/32 Moscow, Russia}

\offprints{J. Chluba, \\ \email{jchluba@mpa-garching.mpg.de}}
\date{Received / Accepted}

\abstract{The CMB angular temperature fluctuations observed by {\sc
Cobe} and \WMAP enable us to place a lower limit on the spectral
distortions of the CMB at any angular scale. These distortions are
connected with the simple fact, that the superposition of blackbodies
with different temperatures in general is not a blackbody. We show,
that in the limit of small temperature fluctuations, the superposition
of blackbodies leads to a y-type spectral distortion.
%------------------------------------------------------------
It is known, that the CMB dipole induces a y-type spectral distortion
with quadrupole and monopole angular distribution leading to a
corresponding whole sky y-parameter of $\yd=2.6\cdot 10^{-7}$.
%------------------------------------------------------------
We show here that taking the difference of the CMB signal in the
direction of the maximum and minimum of the CMB dipole due to the
superposition of two blackbodies
%with the maximal possible temperature difference on the whole CMB sky 
leads to a spectral distortion with $\yo=12\,\yd=3.1\cdot
10^{-6}$. The amplitude of this distortion can be calculated to the
same precision as the CMB dipole, i.e. $0.3\%$ today. Therefore it may
be used as a source with brightness of several or tens of $\mu\,$K to
{\it cross calibrate} and {\it calibrate} different frequency channels
of CMB surveys with precision of a few tens or hundreds of nK.
%------------------------------
%The value of $\yd$ is 12 times smaller than $\yo$ which we are
%proposing here to use for cross calibration. 
%------------------------------
Furthermore, we show in this work that primordial anisotropies for
multipoles $2\leq l \leq 1000$ also lead to spectral distortions but
with much smaller y-parameter, i.e. $y\sim 10^{-11}-10^{-9}$.
%------------------------------------------------------------
}

\maketitle

\section{Introduction}
The \WMAP spacecraft measured the amplitude of the angular
fluctuations of the cosmic microwave background (CMB) temperature with
extremely high precision on a very broad range of angular scales, from
$12'$ up to the whole sky. These temperature anisotropies and the
existence of the acoustic peaks were predicted already long ago
\citep{Peebles70,Suny70a}, but only now after {\sc Boomerang}, {\sc
Maxima}, {\sc Archeops}, {\sc Wmap} and many ground based experiments
like {\sc Cbi}, {\sc Acbar}, {\sc Vsa}, etc.  
%{\sc Boomerang}http://cmb.phys.cwru.edu/boomerang/}, {\sc
%Maxima}\footnote{http://cosmology.berkeley.edu/group/cmb/} and {\sc
%Wmap}\footnote{http://map.gsfc.nasa.gov/},
we know their precise characteristic angular scales and amplitudes.
With no doubt, only tiny fluctuations of the radiation temperature
field have been observed, but nowadays with a precision to better than
1\% down to degree angular scales.

It is commonly assumed that the spectrum in one direction of the sky
is planckian and that only the temperature changes from point to
point. This follows from the nature of the main effects leading to the
appearance of these fluctuations, i.e. the Sachs-Wolfe-effect
\citep{Sachs67} and the Doppler effect due to Thomson scattering off
moving electrons \citep{Suny70a} at redshift $\sim 1100$.  However, as
will be demonstrated below, there are spectral distortions in second
order of $\Delta T/T$.
%------------------------------------------
These distortions are inevitable when the CMB is observed with finite
angular resolution or when regions on the sky containing blackbodies
with different temperatures are compared with each other.

%In the first case due to the average of the
%temperature distribution inside the beam one is in fact observing
%blackbodies with slightly different temperatures simultaneously.
%----------------------
The CMB missions mentioned above have shown that there are
fluctuations of the radiation temperature on the level of $\Delta T
\sim \mu$K$-$mK over a broad range of angular scales.
%----------------------
One may distinguish two basic observational strategies:
%$\mathcal{A}$ a 
(i) {\it absolute} measurements, where the beam flux in some direction
on the sky is compared to an internal calibrator ({\sc Cobe/Firas})
and
%$\mathcal{B}$ an 
(ii) {\it differential} measurements, where the beam flux in one
direction on the sky is compared to the beam flux in another
direction
%\footnote{Usually the angle between the two beams is fixed for differential measurements.}  
({\sc Cobe/Dmr} or {\sc Wmap}).
%------------------------------------------
In the first strategy one observes a sum of blackbodies (\defSB) due
to the average over the beam temperature distribution,
%and compares it to the internal calibrator
whereas in the second two sums of blackbodies are compared with each
other. 
%Therefore, in general in CMB measurements one is dealing with the
%superposition of blackbodies, i.e. the sum and difference of
%blackbodies with different temperatures.
Under these circumstances we will in general speak about the
superposition of blackbodies, i.e. the sum and difference of
blackbodies with different temperatures.
%In this work the spectral distortions arising in both strategies will
%be discuss in detail.

Any experiment trying to extent the great success of the {\sc
Cobe/Firas} instrument, which placed strict upper limits
\citep{Fixsen96,Fixsen02} on a possible $\mu$- \citep{Suny70b},
$|\mu|< 9\cdot 10^{-5}$, and y-type \citep{Zeld69}, $|y|< 1.5\cdot
10^{-5}$, CMB spectral distortion, will only have a finite angular
resolution and would therefore observe a superposition of several
Planck spectra with different temperatures corresponding to the maxima
and minima on the CMB sky as measured with {\sc Wmap}. 

%---------------------------------------------------------------
It is known \citep{Zeld72}, that in the case of a gaussian temperature
distribution this will lead to a spectral distortion indistinguishable
from a y-distortion, with corresponding y-parameter which is
proportional to the dispersion of the temperature distribution. Since
the temperature fluctuations of the CMB indeed are gaussian, this
implies, that the corresponding spectral distortions averaged over
large pieces of the sky should be of y-type. But here we are
interested in the case of measurements with angular resolution of a
few arcminutes to degrees. In this situation, we deal with the limited
statistics of finite regions with different mean temperatures and
therefore it is not obvious, what type of spectral distortion would be
induced in each small patch of the sky.
%----------------------------------
As mentioned above, a similar situation arises when we compare the
signals from two regions on the sky, i.e. take the difference of the
intensities as is usually done in differential observations.
%----------------------------------
Below it will be shown, that for any observation of the CMB
temperature fluctuations, there will be {\it unavoidable spectral
distortions} due to the difference in the temperature of the radiation
we measure and compare
%--------------------------------------------------
and that these distortions will be {\it indistinguishable from a
y-type-distortion}. The biggest distortions arise due to the CMB
dipole.

The unprecedented high sensitivity of future or proposed space
missions like {\sc Planck} and {\sc Cmbpol} or ground based
instruments under construction like {\sc Apex}, the South Pole
Telescope ({\sc Spt}), the Atacama Cosmology Telescope ({\sc Act}) and
{\sc Quest} will offer ways to investigate tiny secondary CMB angular
and spectral fluctuations and should therefore add a lot to the
success of previous missions.
%---------------------------
One target will be the measurement of the SZ effect from clusters,
proto clusters or groups of galaxies and superclusters \citep{Suny72}
or signatures from the first stars in the universe \citep{Oh03}. In
the future CMB experiments will be so sensitive that it will be
possible to investigate in detail the imprints of reionization and the
traces of energy release in the early universe.

\citet{Kaustuv2003} proposed a method to constrain the ionization
history of the universe and the history of heavy element production
using the properties of resonant scattering of CMB photons in the fine
structure lines of oxygen, carbon and nitrogen atoms and ions produced
by the first generation of stars. The strong frequency dependence of
this effect permits to extract the undisturbed angular dependence of
the frequency independent primary temperature fluctuations and thereby
avoid cosmic variance. By comparing the signals in different frequency
channels it is possible to investigate the contributions of the lines
of different ions at different redshifts and therefore to examine
different scenarios of element production and ionization histories in
the low density regions of the universe, with overdensities less than
$\lesssim 10^4$. The sensitivities of {\sc Planck} and {\sc Act}
should be sufficient to detect the signals imprinted by the effects of
resonant scattering, but
%------------------------------------------------------
%it is necessary to {\it cross calibrate} the different frequency
%channels with extremely high accuracy.
%------------------------------------------------------
the crucial point for the successful measurement of {\it any} small
frequency dependent signal is the {\it cross calibration} of the
different frequency channels down to the limits set by the sensitivity
of the experiments. Full sky missions like {\sc Cobe/Dmr} or {\sc
Wmap} normally use the CMB dipole and its annual modulation to check
the calibration of their instruments down to a level of $\mu$K,
whereas experiments with partial sky coverage like {\sc Boomerang}
%which still cover significant fractions of the sky 
are directly using the CMB dipole for calibration issues
\citep{Boomerang}. But both methods permit to cross calibrate
different frequency channels only in first approximation assuming that
the dipole has a planckian spectrum and the same amplitude at all
frequencies.
%------------------------------------------
Unfortunately, this precision of the cross calibration will not be
sufficient to detect the signals from the dark ages as discussed by
\citet{Kaustuv2003}.

%-------------------------------------------------------------------
It is known, that the motion system relative to the CMB restframe in
addition to the dipole generates (in second order of $v/c$) a small
monopole and quadrupole contribution to the CMB brightness of the sky
in the restframe of the observer \citep{Sun80, Bern1990,
Bott1992}. \citet{Sun80}, when they were discussing the radiation
field inside a cluster of galaxies moving relative to the CMB
restframe, have shown that the corresponding dipole induced quadrupole
has a non planckian spectrum, which was then later derived by
\citet{Saz1999}. \citet{Kam03} later applied this solution to the case
of our motion relative to the CMB restframe and proposed to use the
dipole induced quadrupole for calibration purposes.

The solution of \citet{Saz1999} is valid in the case of a narrow beam
observations. In this paper we choose an independent approach, which
is based on the superposition of blackbody spectra with different
temperatures, to look for the maximal and minimal spectral distortion
obtainable from CMB maps. Our method allows us to calculate the value
of the y-parameter for the dipole induced monopole and quadrupole for
a beam with finite width or equivalently for any average of the signal
over extended regions on the sky. Most importantly we show that the
difference of the sky brightness in the direction of the maximum and
minimum of the CMB dipole, corresponding to the maximal difference of
the radiation temperature on the CMB sky, leads to a y-type spectral
distortion with associated y-parameter of $\yo=3.1\cdot 10^{-6}$. 
%-----------------------------------
We propose here to use this spectral distortion arising due to the CMB
dipole to cross calibrate the frequency channels of a CMB experiment
in principle down to the level of a few tens of nK. We discuss
different observing strategies in order to maximize the inferred
spectral distortion (Sect.  \ref{sec:strat}).

%-------------------------------------------------------------------
In this paper, we first give a short summary of the basic equations
necessary in the following derivations and define some of the
terminology used (Sect.~\ref{sec:ingr}). We then discuss the
underlying theory for small spectral distortions
(Sect.~\ref{app:Theorie}) and show that in this limit even the
distortions arising due to the superposition of two blackbodies
(Sect.~\ref{sec:Sup2P}) with close temperatures are well described by
a y-type solution.
%(Sect.  \ref{sec:y}). 
Furthermore, we discuss in detail the spectral distortion due to the
superposition of Planck spectra with different temperatures arising
from to the CMB dipole (Sect.~\ref{app:dipole}) and from the higher
multipoles (Sect.~\ref{sec:yWMAP}) using generated CMB sky maps for
the \WMAP best fit model. We discuss the spectral distortions arising
in differential measurements of the CMB temperature fluctuation
(Sect.~\ref{sec:strat}) and ways to use the spectral distortions
induced by the CMB dipole to cross calibrate the frequency channels of
CMB experiments (Sect.~\ref{sec:Cali}). We end this work with a
discussion of the consequences of the obtained results for some of the
highly demanding tasks which may be addressed by future CMB projects
(Sect.~\ref{sec:conseq}) and finally conclude in Sect.~\ref{sec:conc}.

\section{Basic ingredients}
\label{sec:ingr}
%In this Sect. we give a short summary of the basic equations necessary
%in the following derivations.

\subsection{Compton y-distortion}
\label{sec:y}
In the non relativistic limit, the comptonization of the CMB photons
by hot, isotropic, thermal electrons with Compton y-parameter
%-----------
\beal
\label{eq:y_Param}
y=\int\frac{\kB\Te}{\me c^2}\,\sigT\,\Ne\id l
\Abst{,}
\end{align}
%-----------
where $\Te$ is the temperature of the electron gas, $\sigT$ is the
Thomson cross section and $\Ne$ is the electron number density, leads
to a y-distortion \citep{Zeld69}:
%-----------
\beal
\label{eq:y_Comp}
\frac{\Delta I}{I_0}=y\,\frac{\xn\,e^{\xn}}{e^{\xn}-1}\cdot \left[\xn\,\frac{e^{\xn}+1}{e^{\xn}-1}-4\right]
%\Abst{,}
\end{align}
%-----------
for $y\ll 1$. Here $\xn=h\nu/\kB T_0$ is the dimensionless frequency,
$\nu$ is the photon frequency, $T_0$ is the temperature of the
incoming radiation and $\Delta I=I-I_0$ denotes the difference between
the observed intensity $I$ and undisturbed CMB blackbody spectrum
$I_0=\Ipl(T_0)$, with
%-----------
\beal
\label{eq:I_pl}
\Ipl(T)=\frac{8\pi h}{c^2}\,\frac{\nu^3}{e^{h\nu/\kB T}-1}
\Abst{.}
\end{align}
%-----------
In the Rayleigh Jeans (RJ) limit, i.e. $\xn\ll 1$, equation
\eqref{eq:y_Comp} simplifies to $\Delta T/T_0|_{\rm RJ}=-2\,y$. Since
in this limit a blackbody of temperature $T$ is approximately given by
$B_{\nu, \rm RJ}= \frac{8\pi}{c^2}\,\nu^2\,\kB T$, it is convenient to
compare the comptonized spectrum to a blackbody of temperature $T_{\rm
RJ}=T_0\,[1-2\,y]$, making the relative difference \eqref{eq:y_Comp}
vanish at small $\nu$:
%------------
\beal
\label{eq:y_Dist}
\frac{\Delta I}{\bar{I}}=y\,\frac{\x\,e^{\x}}{e^{\x}-1}\cdot\left[\x\,\frac{e^{\x}+1}{e^{\x}-1}-2\right]
\Abst{.}
\end{align}
%-----------
Here we now defined $\x=h\nu/\kB \TRJ$ and $\bar{I}=\Ipl(T_{\rm
RJ})$ is a blackbody of temperature $\TRJ$.  
%---------------------------------------------
The corresponding difference in the radiation temperature can be
written as
%------------
\beal
\label{eq:DT_T_y}
\frac{\Delta T}{T}=y\cdot g_y(\x)\equiv y\cdot\left[\x\,\frac{e^{\x}+1}{e^{\x}-1}-2\right]
\Abst{.}
\end{align}
%-----------
This equation shows, that the most important characteristic of a
y-distortion is given by the function
%-----------
\beal
\label{eq:g}
g(\x)=\x\,\frac{e^{\x}+1}{e^{\x}-1}\equiv \x \coth \frac{\x}{2} 
\end{align}
%-----------
Whenever the function $g(\x)$ characterizes the frequency dependent
part of the relative temperature difference we may speak about a {\it
y-type} spectral distortion.

\subsection{Relation between temperature and intensity}
\label{sec:T_2_I}
To obtain the relative difference in temperature \eqref{eq:DT_T_y}
corresponding to a y-distortion from the relative difference in
intensity \eqref{eq:y_Dist} the relation 
%------------
\beal
\label{eq:DT_DI}
\frac{\Delta T}{T}=\frac{\Delta I}{I}\,\Abl{\ln T}{\ln
I}=\frac{e^{\x}-1}{\x\, e^{\x}}\,\frac{\Delta I}{I}
%\Abst{,}
\end{align}
%-----------
was implicitly used. In CMB measurement this relation is commonly
applied to {\it relate} intensity differences to temperature
differences, assuming that the difference in intensity obeys
$\frac{\Delta I}{I}\ll 1$ or equivalently $\frac{\Delta T}{T}\ll 1$.
%----------------------------
The sensitivity of future CMB experiments will be extremely high in a
broad range of frequencies and angular scales. Therefore it is
important to understand what corrections arise in next order of
$\frac{\Delta I}{I}$ and $\frac{\Delta T}{T}$ and what kind of
distortions are introduced by using relation \eqref{eq:DT_DI}.
%-----------------------------
For this we consider the simplest case, when we are comparing two pure
blackbodies with different temperatures $T$ and $T'$ and corresponding
intensities $I=\Ipl(T)$ and $I'=\Ipl(T')$. Then using \eqref{eq:I_pl}
the relative difference in intensities can be written as
%-----------
\beal
\label{eq:I_Iref}
\frac{\Delta I}{I}=\frac{I'-I}{I}=\frac{e^\x-1}{e^{\x/(1+\Delta)}-1}-1
\Abst{,}
\end{align}
%-----------
where we defined $\x=h\nu/\kB T$ and $\Delta=(T'-T)/T$. 

Now, using Taylor expansion up to second order in $\Delta$ from
equation \eqref{eq:I_Iref} we obtain
%-----------
\beal
\label{eq:DI_I_appr}
\frac{\Delta I}{I}
&\approx\frac{\x\,e^\x}{e^\x-1}\cdot\left[\Delta
+ g_y(\x)\cdot\frac{\Delta^2}{2}\right]+\mathcal{O}(\Delta^3)
\Abst{,}
\end{align}
%-----------
where $g_y(\x)$ is defined by equation \eqref{eq:DT_T_y}. Now we apply
the relation \eqref{eq:DT_DI} to {\it infer} the relative temperature
difference
%-----------
\beal
\label{eq:DT_T_1}
\frac{\Delta T(\x)}{T}=\Delta+
g_y(\x)\cdot\frac{\Delta^2}{2}+\mathcal{O}(\Delta^3) \Abst{.}
\end{align}
%-----------
This result clearly shows that in second order of $\Delta$ a
y-distortion with y-parameter $y=\Delta^2/2$ is introduced. This
distortion vanishes at low frequencies ($\x\rightarrow 0$) and
increases like $g_y(\x)\sim \x$ in the Wien region ($\x \gg 1$).
%-------------------------------------
Writing $\Delta + g_y(\x)\cdot\Delta^2/2
=\Delta\cdot[1+g_y(\x)\cdot\Delta/2]$ this implies that for {\it any}
given $\Delta$ there is a critical frequency $\x_y$ which fulfills
%---------------
\begin{figure}
\centering
%\resizebox{\hsize}{!}{\includegraphics[width=8.5cm]{./eps/yplot_1.eps}}
\includegraphics[width=\plotwd]
%\plotone
%{./eps/xy.eps}
{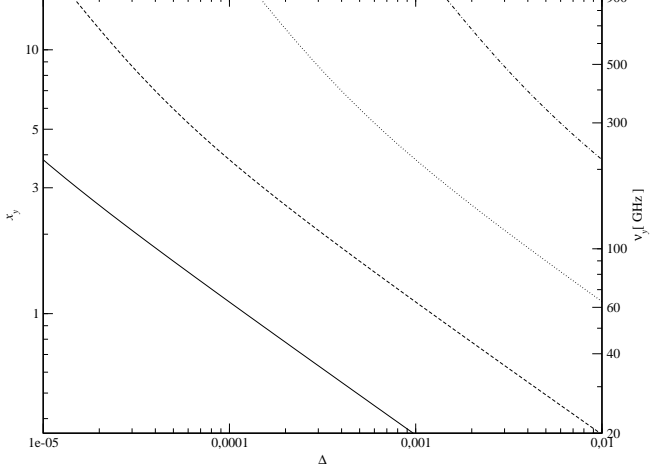}
\caption{$\x_y$ as a function of $\Delta$ for different values of
$\epsilon$ according to equation \eqref{eq:x_y}: For
$\epsilon=10^{-5}$ (solid), $\epsilon=10^{-4}$ (dashed),
$\epsilon=10^{-3}$ (dotted) and $\epsilon=10^{-2}$
(dashed-dotted). The right ordinate corresponds to $\nu_y=\kB
T_0\cdot\x_y/h$, with $T_0=2.725\,$K.}
\label{fig:xy}
\end{figure}
%---------------
%-----------
\beal
\label{eq:x_y}
g_y(\x_y)=\frac{2\,\epsilon}{|\Delta|}
%\Abst{,}
\end{align}
%-----------
and above which the contribution of the y-distortion relative to the
first order term becomes bigger than $\epsilon$. The value of
$\epsilon$ is directly related to the sensitivity of the
experiments. In Fig. \ref{fig:xy} $\x_y$ has been calculated for
different $\epsilon$. At frequencies lower than $\x_y$ the relation
\eqref{eq:DT_DI} is exact within the sensitivity of the experiment,
whereas for $\x>\x_y$ the second order correction has to be taken into
account. In the regime $\x_y\gtrsim 4$ for estimates the simple
approximation $\x_y=2\,\epsilon/|\Delta|+2$ may be used.

%---------------------------------
To give an example, the biggest temperature fluctuation on the whole CMB
sky is due to the CMB dipole. Comparing the maximum and the minimum of
the dipole corresponds to $\Delta\sim 2.48\cdot 10^{-3}$. If some CMB
experiment is able to accurately measure temperature fluctuations on
the level of $\mu$K, contributions of $\epsilon\sim 10^{-4}$ relative
to the dipole signal can be distinguished leading to $\x_{y, \rm
d}\sim 1$. This estimate shows, that at high frequencies spectral
distortions introduced by the usage of the relation \eqref{eq:DT_DI}
for the dipole anisotropy should be taken into account in future CMB
missions like {\sc Planck} and {\sc Act}. In Sect. \ref{app:dipole} we
will discuss these distortions due to the dipole in more detail.

Let us note here, that the temperature difference should not be larger
than a few percent of the temperature of the reference
blackbody. Otherwise corrections due to higher orders in $\Delta$ will
become important and lead to additional distortions (see
Sect. \ref{app:Theorie}). Fortunately, the temperature differences on
the CMB sky are sufficiently small to neglect these corrections.

Equation \eqref{eq:DT_T_1} also shows, that in general the inferred
temperature difference is frequency dependent. At frequencies below
$\x_y$ the {\it inferred} temperature difference is close to the {\it
true} temperature difference $\Delta$ and frequency independent within
the sensitivity of the experiment. In this case, we will speak about a
{\it temperature} distortion or fluctuation, emphasizing that it is
frequency independent. It is possible to eliminate the temperature
distortion using multifrequency measurements. For $\x>\x_y$ frequency
dependent terms become important, which we will henceforth call {\it
spectral} distortions.

\section{Small spectral distortions due to the superposition of blackbodies}  
\label{app:Theorie}
%-----------
When the CMB sky is observed with finite angular resolution or
equivalently if the brightness of parts of the sky (not necessarily
connected) are averaged one is dealing with the sum and more generally
with the superposition of blackbodies. Here we develop a general
formalism to calculate the spectral distortions arising for arbitrary
temperature distribution functions in the limit of small temperature
fluctuations and derive criteria for the applicability of this
approximation.
%-------------------------------------
We first discuss the basic equations necessary to describe the
spectrum of the sum of blackbodies (\defSB) as compared to some
arbitrary {\it reference} blackbody (Sect. \ref{sec:SumP}) and then
generalize these results to the superposition of blackbodies
(Sect. \ref{sec:SupP}).

\subsection{Sum of blackbodies}
\label{sec:SumP}
\begin{figure}
\centering
%\resizebox{\hsize}{!}{\includegraphics[width=8.5cm]{./eps/yplot_1.eps}}
\includegraphics[width=\plotwd]
%\plotone
{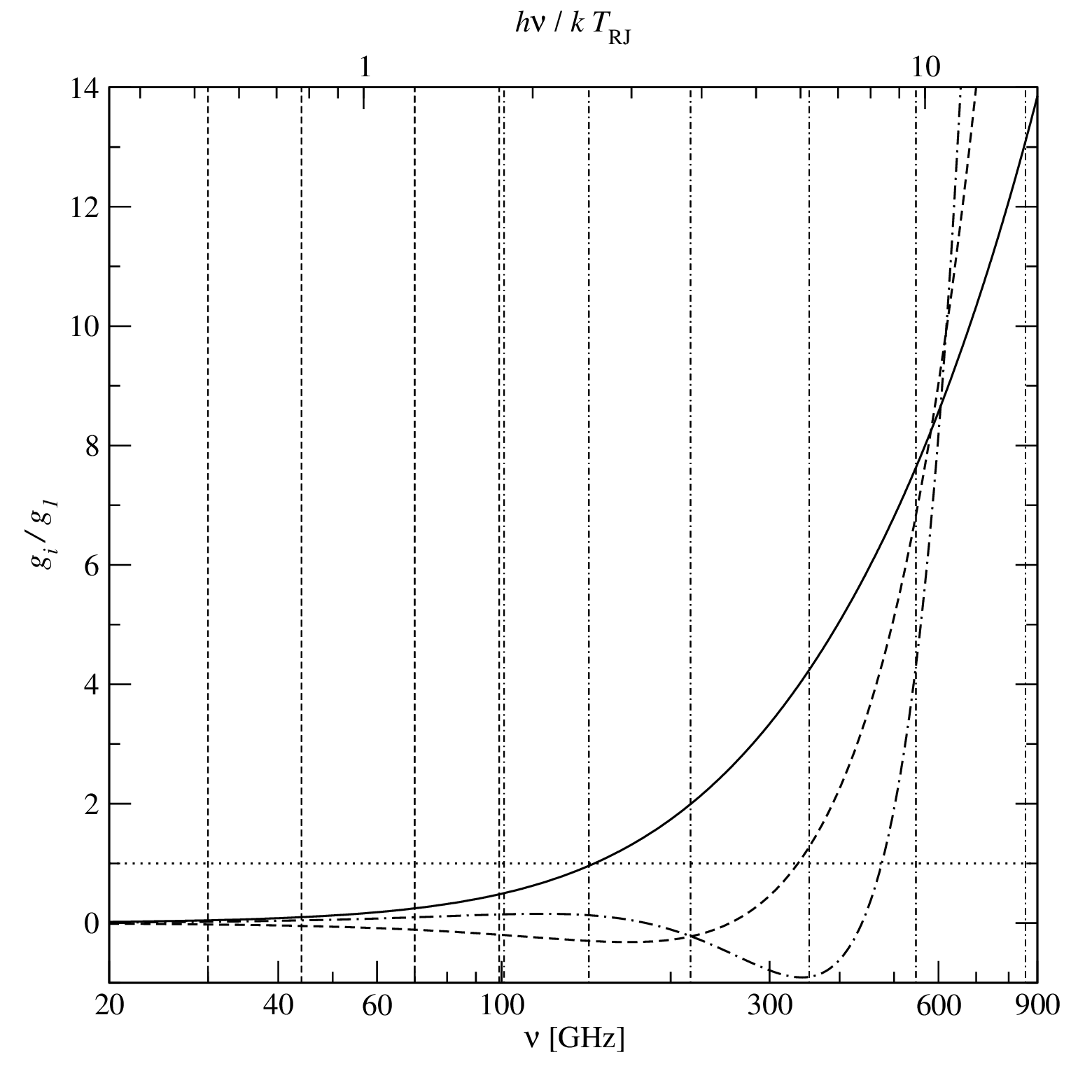}
\caption{Frequency dependence of $g_i/g_1\equiv\Delta T_i/T_{\rm
RJ}\,\delta^i$: Dotted line $\Delta T_1/\TRJ\,\delta$, which is
equivalent to a temperature distortion, solid line $g_y(\x)=2\Delta
T_2/\TRJ\,\delta^2$, which is equivalent to a y-distortion $\Delta
T/y\,T_0$ as given by equation \eqref{eq:DT_T_y}, dashed line $\Delta
T_3/\TRJ\,\delta^3$ and dashed-dotted line $\Delta
T_4/\TRJ\,\delta^4$. The function $g_y(\x)$ becomes larger than unity
for $\nu\geq 146.2\,$ GHz or $\x\geq 2.57$, larger than 5 for $\nu\geq
396.7\,$GHz or $\x\geq 6.99$ and larger than 10 for $\nu\geq
681.3\,$GHz or $\x\geq 12$. For $\nu= 100\,$ GHz or $\x= 1.76$ it
follows $g_y= 0.5$. Also shown as vertical lines are the {\sc Planck}
LFI (dashed) and HFI (dash-dotted) frequency channels.}
\label{fig:four}
\end{figure}

Following the paper of \cite{Zeld72} we express the total spectrum
resulting for the SB with different temperatures $T$ as:
%-----------
\beal
\label{eq:nt}
I(\nu)=\int R(T)\,\Ipl(T)\id T \Abst{.}
\end{align}
%-----------
Here $R(T)$ denotes the normalized ($\int R\id T=1$, $R\geq 0$)
temperature distribution function, which will be used below to model
the {\it beam} of some CMB experiment, and $\Ipl(T)$ is a Planck
spectrum of temperature $T$ as given by equation \eqref{eq:I_pl}. 
%-----------
Now we want to compare the spectrum $I(\nu)$ to a {\it reference}
blackbody of temperature $\Tref$. Defining $\delta=(T-T_{\rm
ref})/\Tref$ and inserting $T=\Tref\,(1+\delta)$ into
equation \eqref{eq:I_pl}, $\Ipl$ may be rewritten as
%-----------
\beal
\label{eq:I_pl_x}
\Ipl(\delta)=A\,\Tref^3\,\frac{\x^3}{e^{\x/(1+\delta)}-1}
\end{align}
%-----------
where $\x=h\nu/\kB \Tref$ is the dimensionless frequency and
$A=8\pi\kB^3/h^2\,c^2$.
%-----------
CMB temperature fluctuations are of the order $\Delta T/T\sim
10^{-5}-10^{-3}$. Therefore we are interested in the case when $|\delta|\ll 1$. This
allows us to perform a Taylor expansion of equation \eqref{eq:I_pl_x}:
%-----------
\bmulti
\label{eq:Ipl_Taylor}
\Ipl(\delta)=\left.\Ipl\right|_{\delta=0}\\
-\delta\cdot \x\left.\pAb{\Ipl}{\xi}\right|_{\delta=0}
+\frac{\delta^2}{2}\cdot \x\,\left[\x\,\PAb{\Ipl}{\xi}{2}
%\right.
%\\
%\left.
+2\,\pAb{\Ipl}{\xi}\right]_{\delta=0}\\
-\frac{\delta^3}{6}\cdot \x\,\left[\x^2\,\PAb{\Ipl}{\xi}{3}
+6\,\x\,\PAb{\Ipl}{\xi}{2}+6\,\pAb{\Ipl}{\xi}\right]_{\delta=0}
\\
+\frac{\delta^4}{24}\cdot \x\,\left[\x^3\,\PAb{\Ipl}{\xi}{4}
+12\,\x^2\,\PAb{\Ipl}{\xi}{3}
\right.
\\
\left.
+36\,\x\,\PAb{\Ipl}{\xi}{2}+24\,\pAb{\Ipl}{\xi}\right]_{\delta=0}
+\mathcal{O}(\delta^5)
\Abst{,}
\end{multline}
%-----------
where here $\xi=\x/(1+\delta)$ was defined. Keeping only terms up to
fourth order in $\delta$ this simplifies to:
%-----------
\beal
\label{eq:Ipl_app}
\Ipl(\delta)=\Ir\,\sum^{4}_{k=0} g_k(\x)\,\delta^k
\Abst{.}
\end{align}
%-----------
Here we used the abbreviation $\Ir=\Ipl(\delta=0)$ for the
reference blackbody spectrum of temperature $\Tref$ and defined
the functions $g_k(\x)$ as
%-----------
\bsub
\label{eq:g_k}
\beal
g_0(\x)&=1\\
g_1(\x)&=\frac{\x\,e^{\x}}{e^{\x}-1}\\
g_2(\x)&=\frac{g_1}{2}\left[\x\,\frac{e^{\x}+1}{e^{\x}-1}-2\right]\equiv
\frac{g_1}{2}\cdot g_y(\x)\\
g_3(\x)&=g_1\left[\frac{\x^2}{6}\,\frac{\cosh{\x}+2}{\cosh{\x}-1}-\frac{2 g_2}{g_1}-1\right]\\
g_4(\x)&=g_1\left[\frac{\x^3}{24}\,\frac{6\coth{\frac{\x}{2}}+\sinh{\x}}{\cosh{\x}-1}
-3\,\frac{g_2+g_3}{g_1}-1\right]
%-\frac{3\,(g_2+g_3)+g_1}{g_1}\right]
%\Abst{,}
\end{align}
\esub
%-----------
where $g_y(\x)$ is defined by equation \eqref{eq:DT_T_y}. In
Fig. \ref{fig:four} the frequency dependence of the functions
$g_i(\x)$ is illustrated. Deviations from a blackbody spectrum become
most important in the Wien region and vanish in the RJ region of the
CMB spectrum.
%---------------
%---------------

Inserting equation \eqref{eq:Ipl_app} into \eqref{eq:nt} the relative
difference $\Delta I/\Ir=(I-\Ir)/\Ir$ between the SB
spectrum and the reference blackbody $\Ir$ can be derived
%-----------
\beal
\label{eq:nt_gen}
\frac{\Delta I}{\Ir}\approx\sum^{4}_{k=1} g_k(\x)\,\left<\delta^k\right>
\Abst{,}
\end{align}
%-----------
where we introduced the abbreviation $\left<\delta^k\right>=\int
R\,\delta^k\id T$ for the $k$-th moment of the temperature
distribution. To find the corresponding difference in temperatures
from \eqref{eq:nt_gen} one may use the relation \eqref{eq:DT_DI}. As
mentioned before the {\it inferred} temperature difference in general
will be frequency dependent.

%-----------
In equation \eqref{eq:nt_gen} the term proportional to the first
moment, $\left<\delta^1\right>$, corresponds to a temperature
distortion which introduces a frequency independent shift in the
measured temperature difference. It is possible to eliminate this
contribution by multifrequency measurements, since it does not change
with frequency. The term proportional to the second moment is
indistinguishable from a Compton y-distortion as given by equation
\eqref{eq:y_Dist} with y-parameter
%-----------
\beal
\label{eq:y_Sup}
y_{\rm S}=
%{\textstyle \frac{1}{2}}\,\left<\delta^2\right>
\frac{\left<\delta^2\right>}{2} 
\Abst{.}  
\end{align}
%-----------
Higher moments lead to additional spectral distortions making the
total spectral distortion differ from a pure y-distortion. As will be
shown below for the CMB these higher order corrections can be
neglected.

\subsection*{Limiting cases for $\frac{\Delta T}{\Tref}$}
In the RJ limit ($\x\ll 1$), starting from \eqref{eq:nt} and using
relation \eqref{eq:DT_DI} one can find
%-----------
\bsub
\label{eq:nt_RJ}
\beal
\label{eq:nt_RJa}
\left.\frac{\Delta T}{\Tref}\right|_{\rm RJ}
&\stackrel{\stackrel{\x\ll 1}{\downarrow}}{\approx}
\left<\delta\right>+\frac{\x^2}{12}\left<\frac{\delta^2}{1+\delta}\right>
\\
\label{eq:nt_RJb}
&\!\!\!\!\!
\stackrel{\stackrel{\x\ll 1\wedge\delta\ll 1}{\downarrow}}{\approx}
\left<\delta\right>
+\frac{\x^2}{12}\Big[\left<\delta^2\right>-\left<\delta^3\right>
+\left<\delta^4\right>\Big]
\Abst{.}
\end{align}
\esub
%-----------
This result shows, that for $\x\rightarrow 0$ the functions $g_2$,
$g_3$, and $g_4$ all vanish like $\sim x^2$, reflecting the fact that
the sum of RJ spectra is again a RJ spectrum and that no spectral
distortions are expected for $\x\ll 1$. In the first step we have only
assumed that $\x\ll 1$. Therefore, up to second order in $\x$ equation
\eqref{eq:nt_RJa} describes the SB for any $\delta$ to very high
accuracy. For $\delta\gg 1$ equation \eqref{eq:nt_RJa} takes the form
$\Delta T/\Tref|_{\rm RJ}\approx
\left<\delta\right>\cdot[1+\frac{\x^2}{12}]$.
%------------------------------------------------
For the approximation \eqref{eq:nt_RJb} we have in addition to $\x
\ll1$ assumed that $\delta\ll 1$. In this limit the same result can be
obtained starting from equation \eqref{eq:nt_gen}.

%-----------
In the Wien region of the CMB spectrum ($\x\gg 1$) again starting from
\eqref{eq:nt} and using relation \eqref{eq:DT_DI} one can deduce
%-----------
\bsub
\label{eq:nt_Wien}
\beal
\label{eq:nt_Wiena}
\left.\frac{\Delta T}{\Tref}\right|_{\rm W}
&\stackrel{\stackrel{\x\gg 1}{\downarrow}}{\approx}
\frac{1}{x}\cdot\left[\left<\exp\left(x\cdot\frac{\delta}{1+\delta}\right)\right>-1\right]
%=\sum_{k=1}^\infty \frac{x^{k-1}}{k!}\cdot\left<\left(\frac{\delta}{1+\delta}\right)^k\right>
\\
\label{eq:nt_Wienb}
&\!\!\!\!\!\!\!
\stackrel{\stackrel{\x\gg 1\wedge x\cdot\delta\ll 1}{\downarrow}}{\approx}
\left<\delta\right>-\left<\delta^2\right>+\left<\delta^3\right>-\left<\delta^4\right>
\nonumber\\
&\qquad\quad
+\frac{\x}{2}\left[\left<\delta^2\right>
+\frac{\x}{3}\,\left<\delta^3\right>
+\frac{\x^2}{12}\,\left<\delta^4\right>\right]
\Abst{.}
\end{align}
\esub
%-----------
Equation \eqref{eq:nt_Wiena} is valid for $\x\gg 1$ and arbitrary
$\delta$, whereas \eqref{eq:nt_Wienb} only is applicable for $\x\gg 1$
and $x\cdot\delta\ll 1$. Result \eqref{eq:nt_Wienb} can be also
obtained starting from equation \eqref{eq:nt_gen}.

\subsection*{When higher order moments are important}
Using multifrequency observations it is possible to eliminate the
temperature distortion ($\propto \left<\delta\right>$). In this case,
the leading term in the signal corresponds to the y-distortion. Here
we are interested in the question when higher order moments start to
contribute significantly to the total spectral distortions.
%-----------
In the RJ region using equation \eqref{eq:nt_RJb} one can find that
for
%-----------
\beal
\label{eq:crit_RJ}
\epsilon\cdot\left<\delta^2\right> > |\left<\delta^3\right>-\left<\delta^4\right>|
%\Abst{,}
\end{align}
%-----------
the relative contributions of the higher moments to the y-distortion
is less than $\epsilon$. 
%----------------------------------------------------------
This equation shows, that in general corrections due to the higher
moments will be of the order of $\sim|\left<\delta^3\right>|/\left<\delta^2\right>$. For
temperature distributions which are symmetric around the reference
temperature all the odd moments vanish. In this case the corrections
typically will be of the order of $\sim\left<\delta^4\right>/\left<\delta^2\right>$.
%----------------------------------------
If we consider the simple case, when we compare two pure blackbodies
with temperatures $T_1$ and $T_2$ and use $\Tref=T_1$ then the moments
are given as $\left<\delta^k\right>=\Delta^k$, with
$\Delta=(T_2-T_1)/T_1$. From this we can conclude, that the relative
contribution of the third moment to the spectral distortion will be
$\lesssim|\Delta|$. Therefore, in the RJ region of the spectrum even
for $\Delta\sim 10^{-2}$ higher order moments will lead to corrections
$\lesssim 1\%$ to the y-distortion. For the CMB with $\Delta\sim
10^{-5}-10^{-3}$ these corrections can be safely neglected.

In the Wien region the situation is a bit more complicated. Given some
value of $\epsilon$ one can find the frequency $\xc$ above which the
relative contribution of higher order moments will become
important. To find $\xc$ one has to solve the non-linear equation
%---------------
\begin{figure}
\centering
%\resizebox{\hsize}{!}{\includegraphics[width=8.5cm]{./eps/yplot_1.eps}}
\includegraphics[width=\plotwd]
%\plotone
{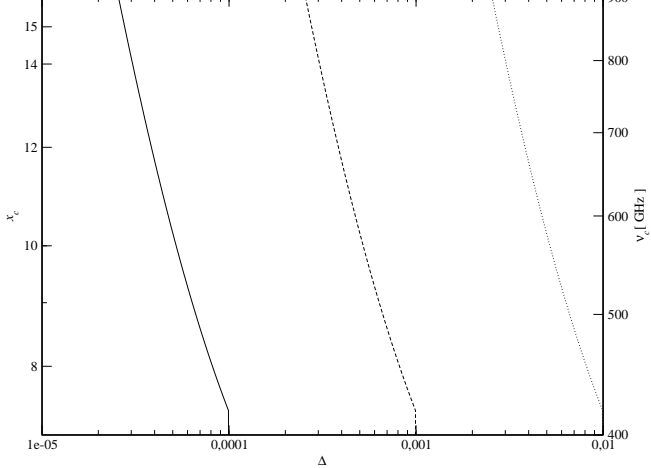}
\caption{$\x_y$ as a function of $\Delta$ for different values of
$\epsilon$ according to equation \eqref{eq:xc_equation} assuming that
$\left<\delta^k\right>=\Delta^k$: For $\epsilon=10^{-4}$ (solid),
$\epsilon=10^{-3}$ (dashed) and $\epsilon=10^{-2}$ (dotted). The right
ordinate corresponds to $\nu_y=\kB T_0\cdot\x_y/h$, with
$T_0=2.725\,$K.}
\label{fig:xc_no_g1}
\end{figure}
%---------------
%-----------
\beal
\label{eq:xc_equation}
\epsilon&=\frac{g_3(\xc)\left<\delta^3 \right>+g_4(\xc)\left<\delta^4\right>}{g_2(\xc)\left<\delta^2\right>}
%\Abst{,}
\end{align}
%-----------
or an equivalent equation directly using \eqref{eq:nt}. 
%----------------------------------------
In Fig. \ref{fig:xc_no_g1} $\xc$ has been calculated for different
$\epsilon$ and $\Delta$ again considering the simple case, when we
compare two pure blackbodies with temperatures $T_1$ and $T_2$ and use
$\Tref=T_1$. Higher order moments contribute 1\% to the total spectral
distortion above $\xc=10$ for $\Delta\sim 5\cdot 10^{-3}$, which is
roughly 5 times the dipole amplitude. For $|\Delta|\lesssim 3\cdot
10^{-3}$ corrections are less than $1\%$ even in the highest {\sc
Planck} frequency channels. In the limit $\xc\gtrsim 6$ for estimates
of the critical frequency above which higher order correction become
bigger than $\epsilon$ one may use
%-----------
\beal
\label{eq:xc}
\xc=\min\left[3\,\epsilon\,\frac{\left<\delta^2\right>}{|\left<\delta^3\right>|}+4, 
\sqrt{12\,\epsilon\,\frac{\left<\delta^2\right>}{\left<\delta^4\right>}}+5\,\right]
\Abst{.}
\end{align}
%-----------

In summary, the above discussion shows that, as long as
$|\Delta|<\epsilon$ the spectral distortion for $\x<\xc$ is given by a
pure y-distortion. In this case the biggest deviations from a
y-distortion are expected in the Wien region of the CMB spectrum,
where the sensitivity of the {\sc Planck} mission and other future
experiments is high. But even for the dipole anisotropy these
corrections should be less than 1\%.
%----------------------------------------------
If $|\Delta|\sim\epsilon$ deviations in the RJ become important and
higher order corrections have to be taken into account. 
%------------------------------------------------------------
Given the amplitude of the temperature fluctuations on the CMB sky,
$\Delta\sim 10^{-5}-10^{-3}$, corrections to the y-distortion due to
higher moments will be less than 1\% in the highest {\sc Planck}
frequency channels even for the dipole amplitude and certainly less
than $0.1\%$ for the higher multipoles. Therefore we can safely
describe the spectral distortions arising due to the superposition of
blackbodies with slightly different temperatures by y-distortions. In
what follows below we assume that higher order corrections are
negligible.

\subsection*{Beam spectral distortion}
In the RJ limit the sum of blackbodies (\defSB) is again a blackbody
with temperature $\TRJ=\int R\,T\id T$. The RJ temperature $\TRJ$ can
be directly measured at sufficiently low frequencies and hence it is
convenient to compare the \SoBs to a {\it reference} blackbody of
temperature $\TRJ$. This makes any distortion due to the SB vanish at
low $\nu$.

In the following, we will call $\left<\delta^k\right>$ {\it beam}
moment, when $\delta=(T-\TRJ)/\TRJ$ is defined as the relative
difference between the temperature $T$ of a given blackbody inside the
beam and the RJ temperature of the SB, i.e. the beam RJ temperature
$\TRJ$. For this choice of reference blackbody the temperature
distortion vanishes and we obtain
%-----------
\beal
\label{eq:nt_appr}
\frac{\Delta I}{\bar{I}}\approx g_2(\x)\,\left<\delta^2\right>
%+g_3(\x)\,\left<\delta^3\right>+g_4(\x)\,\left<\delta^4\right>
\Abst{,}
\end{align}
%-----------
where $\x=h\nu/\kB \TRJ$ and we introduced $\bar{I}=\Ipl(\TRJ)$. We
should mention again that the beam spectral distortion arising due to
the SB with different temperatures can be approximated by a pure
y-distortion with y-parameter
%-----------
$y_{\rm S}$ as given by \eqref{eq:y_Sup},
%%-----------
%\beal
%\label{eq:y_Sup}
%y_{\rm S}=\frac{\left<\delta^2\right>}{2} 
%\Abst{,}  
%\end{align}
%%-----------
if only the second moment of the temperature distribution,
$\left<\delta^2\right>$, is important. In general higher moments have
to be taken into account leading to additional spectral
distortions. 

\subsection*{Sum of two Planck spectra with different weights} 
\label{app:Sum2P_weights}
%------------
%\subsubsection*{Beam spectral distortion} 
%------------
For the sum of two Planck spectra with temperatures $T_1$ and $T_2$ of
relative weights $w_1$ and $w_2=1-w_1$ the RJ temperature is given by
$\bar{T}=w_1\,T_1+w_2\,T_2$. For the first five beam moments one may
obtain:
%-----------
\bsub
\label{eq:moments_two}
\beal
\label{eq:moments_twoa}
\left<\delta^0\right>&=1, \qquad \left<\delta^1\right>=0\\
\label{eq:moments_twob}
\left<\delta^2\right>&=w_1\,w_2\cdot\left[\frac{\Delta T}{\bar{T}}\right]^2\\
\label{eq:moments_twoc}
\left<\delta^3\right>&=w_1\,w_2\,(1-2\,w_1)\cdot\left[\frac{\Delta T}{\bar{T}}\right]^3\\
\label{eq:moments_twod}
\left<\delta^4\right>&=w_1\,w_2\,(1-3\,w_1\,w_2)\cdot\left[\frac{\Delta T}{\bar{T}}\right]^4
\Abst{.}
\end{align}
\esub
%-----------
Here $\Delta T=T_1-T_2\ne 0$ was defined. 
%-----------
Making use of equation \eqref{eq:xc}, in general ($w_1\ne w_2 \ne 0$)
the spectral distortion significantly deviates from a y-distortion by
a fraction $\epsilon$ at frequencies
%-----------
\beal
\label{eq:xc_two}
\x\geq \min\left[\frac{3\,\epsilon}{|1-2\,w_1|}, 
\sqrt{\frac{12\,\epsilon}{1-3\,w_1\,w_2}}\,\right]\cdot\frac{\bar{T}}{|\Delta T|}
\end{align}
%-----------
If one of the weights $w_1$ or $w_2$ vanishes then, as expected, there
is no distortion. In the case $w_1\equiv w_2$, all the odd moments
vanish and the even moments are given by
$\left<\delta^n\right>=\left[\Delta T/(T_1+T_2)\right]^n$. Then the
critical frequency $\xc=2\cdot\sqrt{3}\cdot 10^{-1}\,(T_1+T_2)/|\Delta
T|$ above which the spectral distortion deviates significantly from a
y-distortion becomes 10 only if $|\Delta T|\approx
0.069\,\bar{T}$. This implies that for non gaussian temperature
distributions deviations from a y-distortion will only be important if
the temperature difference becomes of the order of a few $\%$ of the
RJ temperature. This is clearly not the case for the CMB in the
observed universe. Even the dipole anisotropy leads to a y-distortion
up to very high frequencies and we can safely neglect the
contributions of moments of the temperature distribution with $k>2$.

\subsection*{Difference to an arbitrary reference blackbody} 
Here we want to derive the equations describing the comparison of the
spectrum of the SB with an arbitrary reference blackbody. This example
can be applied for {\it absolute} measurements of the CMB. 
%, where the beam flux is compared to the flux of an internal
%calibrator.
%---------------------------------------
Starting again from equation \eqref{eq:nt_gen} and using the results
obtained for the beam spectral distortions, one can rewrite the moments
of $\delta_{\rm ref}=(T-\Tref)/\Tref$ in terms of the {\it beam}
moments, $\left<\delta^k\right>$ and $\DeltaRJ=(\TRJ-T_{\rm
ref})/\Tref$, with $\delta=(T-\TRJ)/\TRJ$: %, leading to
%-----------
\bsub
\label{eq:delref}
\beal
\label{eq:delrefa}
\left<\delta_{\rm ref}^1\right>&=\DeltaRJ
\\
\label{eq:delrefb}
\left<\delta_{\rm ref}^2\right>&=\left<\delta^2\right>+\DeltaRJ^2
+\mathcal{O}\left(\left<\delta^2\right>\cdot\DeltaRJ\right)
\Abst{.}
\end{align}
\esub
%-----------
Using formula \eqref{eq:nt_gen}, one can directly infer the relative
difference between the spectrum of the SB and the {\it chosen}
reference blackbody $I_{\rm ref}=\Ipl(\Tref)$:
%-----------
\beal
\label{eq:nt_appr_ref}
\frac{I-I_{\rm ref}}{I_{\rm ref}}
&=g_1(\x_{\rm ref})\,\DeltaRJ+g_2(\x_{\rm ref})\,\left[\DeltaRJ^2+\left<\delta^2\right>\right]
\Abst{,}
\end{align}
%-----------
where $\x_{\rm ref}=h\nu/\kB \Tref$ and we assumed that the
contributions of $\left<\delta^2\right>\cdot\DeltaRJ$ and any higher
moments of both $\delta$ and $\DeltaRJ$ are negligible. Note that
$g_2(\x_{\rm ref})=g_2(\x)\cdot[1+\mathcal{O}(\DeltaRJ)$] and
therefore $g_2(\x_{\rm ref})\left<\delta^2\right>\equiv\Delta
I/\bar{I}$, where $\Delta I/\bar{I}$ is given by equation
\eqref{eq:nt_appr}.

%---------------------------------
Equation \eqref{eq:nt_appr_ref} shows that the spectral distortion of
the SB spectrum with respect to the some reference blackbody $I_{\rm
ref}$ has the following three contributions:

\begin{enumerate}

\item[(i)] A temperature distortion $\propto\DeltaRJ$. This distortion
vanishes if the SB is compared to a blackbody with temperature $T_{\rm
ref}=\TRJ$. Since the y-distortion vanishes in the RJ region of the
CMB spectrum $\DeltaRJ$ can be directly measured at low frequencies.

\item[(ii)] A y-distortion $\propto\DeltaRJ^2$ due to the dispersion of
the RJ temperature of the SB with respect to the chosen reference
temperature. This contribution also vanishes if the chosen reference
blackbody has temperature $\Tref=\TRJ$.

\item[(iii)] A y-distortion $\propto\left<\delta^2\right>$ due to the
dispersion of the temperature $T$ with respect to the RJ temperature
of the SB. Only if the temperature distribution function is a Dirac
$\delta$-function, i.e. when all the blackbodies inside the beam have
the same temperature or equivalently when the angular resolution of
the experiment is {\it infinite}, this term vanishes. For the SB it
sets the minimal value of the y-parameter for a given temperature
distribution function.

\end{enumerate}

Equation \eqref{eq:nt_appr_ref} represent the most general case for
the comparison of the SB with any reference blackbody in the limit of
small temperature fluctuations around the temperature of the reference
blackbody. It implies that in this limit the y-distortion arising due
to the comparison of the SB spectrum with some chosen reference
blackbody with temperature $\Tref$ is completely determined by the
second beam moment and the beam RJ temperature.

\subsection{Superposition of Planck spectra} 
\label{sec:SupP}
Using the results obtained so far, it is possible to generalized the
sum of Planck spectra to the superposition of Planck spectra. This
case applies for {\it differential} measurements of the CMB, where we
directly compare two identical beams with intensities $I_1$ and $I_2$
and measure the intensity difference $\Delta I=I_2-I_1$. If we want to
apply the relation \eqref{eq:DT_DI} to relate $\Delta I\leftrightarrow
\Delta T$ then we have to fix the {\it reference} temperature $\Tref$
and thereby define the reference blackbody $\Iref$. During the
discussion in Sect. \ref{sec:T_2_I} we used $\Tref=T$ (cf. equation
\eqref{eq:I_Iref}) but in principle one is free to choose any
temperature for which the limit of small spectral distortions is
justified, since $\Delta I=I_2-I_1=I_2-I_{\rm ref}+I_{\rm
ref}-I_1$. In the case of the CMB, one will usually set $\Tref$ equal
to the whole sky mean temperature $T_0$, but as will be shown below
the {\it inferred} y-distortion strongly depends on this choice.

Defining $\Delta_i=(T_i-\Tref)/\Tref$, where $T_i$ are the
temperatures of the blackbodies inside the beam $i$, and applying
formula \eqref{eq:nt_gen} one may obtain
%-----------
\beal
\label{eq:Sup_plancks}
\frac{I_2-I_1}{I_{\rm ref}}
=g_1(\x)\,\left[\Deltas+g_y(\x)\cdot \ydifs\right]
\Abst{,}
\end{align}
%-----------
with $\x=h\,\nu/\kB \Tref$ and 
%-----------
\bsub
\label{eq:D_D2}
\beal
\label{eq:D_D2a}
\Deltas
&=\left<\Delta_2\right>_{\rm b, 2}-\left<\Delta_1\right>_{\rm b, 1}=\frac{\bar{T}_{\rm b,2}-\bar{T}_{\rm b,1}}{\Tref}
\\
\label{eq:D_D2b}
\ydifs&=\frac{1}{2}\left[\left<\Delta^2_2\right>_{\rm b, 2}-\left<\Delta^2_1\right>_{\rm b, 1}\right]
\Abst{,}
\end{align}
\esub
%------------
where $\left<X\right>_{\rm b,\it i}$ denotes the corresponding beam
averages and we defined the beam RJ temperatures $\bar{T}_{\rm b, \it
i}=\left<T_i\right>_{\rm b,\it i}$. In equation \eqref{eq:Sup_plancks}
we have neglected the contributions of any higher moments of
$\Delta_i$. Here the temperature distortion is directly related to the
relative difference of the RJ temperatures.

The y-parameter $\ydifs$ can be rewritten using the beam moments
$\big<\delta_i^2\big>_{\rm b,\it i}$, with $\delta_i=(T_i-\bar{T}_{\rm
b,\it i})/\bar{T}_{\rm b,\it i}$ and defining $\Delta_{\rm ref, \it
i}=\left<\Delta_i\right>_{\rm b, \it i}=(\bar{T}_{\rm b,\it i}-\Tref)/\Tref$:
%-----------
\bmulti
\label{eq:rewrite_the_shit}
\ydifs
%=\left<\Delta^2_2\right>_{\rm b,2}-\left<\Delta^2_1\right>_{\rm b,1}\nonumber\\
=\frac{1}{2}\,\left[\left<\delta_2^2\right>_{\rm b,2}-\left<\delta_1^2\right>_{\rm b,1}\right]
\\
+\frac{1}{2}\,\left[
\Delta^2_{\rm ref,2}-\Delta^2_{\rm ref,1}\right]
+\mathcal{O}(\left<\delta_i^2\right>_{\rm b,\it i}\cdot \Delta_{\rm ref,\it i})
\Abst{.}
\end{multline}
%------------
The difference of the beam moments $\big<\delta_i^2\big>_{\rm b,\it
i}$ does not depend on the chosen reference temperature $\Tref$ and
therefore always leads to the same contribution
%-----------
\beal
\label{eq:ydifb}
\ydifb=\frac{1}{2}\,\left[\left<\delta_2^2\right>_{\rm b,2}-\left<\delta_1^2\right>_{\rm b,1}\right]
\end{align}
%------------
to the inferred y-parameter. In the limit of high angular resolution
it vanishes. For the second term in \eqref{eq:rewrite_the_shit}
($\propto \Delta^2_{\rm ref,2}-\Delta^2_{\rm ref,1}$) one finds
%-----------
\beal
\label{eq:D2_D1}
\yxxx(\Tref)=\frac{1}{2}\,\left[\Delta^2_{\rm ref,2}-\Delta^2_{\rm ref,1}\right]
=\Deltas\cdot\DeltaRJ
\Abst{,}
\end{align}
%------------
with $\DeltaRJ=(\TRJ-\Tref)/\Tref$ and $\TRJ=(\bar{T}_{\rm
b,1}+\bar{T}_{\rm b,2})/2$. This contribution to the total inferred
y-parameter, 
$\ydifs(\Tref)=\ydifb+\yxxx(\Tref)$, 
%%-----------
%\beal
%\label{eq:y_total}
%\ydifs(\Tref)=\ydifb+\yxxx(\Tref)
%\Abst{,}
%\end{align}
%%------------
depends on the choice of $\Tref$. Here it becomes clear that setting
$\Tref\equiv\TRJ$, i.e. to the average RJ temperature of the sum of
{\it both} beams, only the difference of the second {\it beam} moments
leads to a y-distortion. Let us note here, that equation
\eqref{eq:nt_appr_ref} is a special case of \eqref{eq:Sup_plancks},
where beam 1 has no contribution of the beam moment
$\left<\delta_1^2\right>_{\rm b,1}$, i.e. where the beam spectrum is a
blackbody with temperature $\bar{T}_{\rm b,1}$, and where
$\Tref=\bar{T}_{\rm b,1}$.

\subsection*{Dependence on the reference temperature} 
Here we want to clarify some aspects of the dependence of the inferred
y-parameter on the chosen reference temperature. If we assume that
$\ydifb\ll 1$ and $\bar{T}_{\rm b,2}>\bar{T}_{\rm b,1}$, we can find
the reference temperature at which the total y-distortion vanishes:
%-----------
\beal
\label{eq:Tref_y_is_0}
T_{y=0}=\frac{\bar{T}_{\rm b,1}+\bar{T}_{\rm b,2}}{2}\,\left[1
+\frac{\bar{T}_{\rm b,1}+\bar{T}_{\rm b,2}}{\bar{T}_{\rm b,2}-\bar{T}_{\rm b,1}}\cdot\frac{\ydifb}{2}\right]
\Abst{.}
\end{align}
%------------
In the limit of high angular resolution ($\yb=0$) we find that
$T_{y=0}=\TRJ$. One encounters this situation in differential
measurements of the CMB brightness between two points on the sky. For
$\bar{T}_{\rm b,2}>\bar{T}_{\rm b,1}$ and $\ydifb>0$ the reference
temperature at which the y-distortion vanishes is always positive and
also the temperature distortion corresponding to the first term in
equation \eqref{eq:Sup_plancks} is positive.
%---------------------------------------------------
Then for the inferred y-distortion the following three regimes can be
defined: (i) for $\Tref < T_{y=0}$ the inferred y-distortion is
positive, (ii) for $\Tref = T_{y=0}$ per definition there is no
y-distortion up to first order in $\ydifb$ and (iii) for
$\Tref>T_{y=0}$ the inferred y-distortion is negative even though
$\bar{T}_{\rm b,2}>\bar{T}_{\rm b,1}$. This shows that for the
superposition of Planck spectra it is in general not sufficient to
choose the beam intensity with higher RJ temperature as $I_2$ in order
to obtain a positive y-parameter.

Now we are interested in the change of $\Deltas$ and $\ydifs$, if we
change from one reference temperature $\Tref$ to another $\Treft$. For
$\Treft$ one can again write down an equation similar to
\eqref{eq:Sup_plancks}, where now $\Delta_i=(T_i-\Tref)/\Tref$ is
replaced by $\Delta'_i=(T_i-\Treft)/\Treft$ and $\x=h\nu/\kB \Tref$ by
$\x'=h\nu/\kB \Treft$. Using \eqref{eq:D_D2} and
\eqref{eq:rewrite_the_shit}, we can write
%-----------
\bsub
\label{eq:D_D2'}
\beal
\label{eq:D_D2a'}
\Deltas'&=\frac{T_2-T_1}{\Treft}=\Deltas \cdot[1+\Delta_{\rm r}]
\\
\label{eq:D_D2b'}
\ydifs'&=\yb'+\yxxx'(\Treft)
\stackrel{\stackrel{\yb'\equiv\yb}{\downarrow}}{=} 
\yb+\Deltas'\cdot\DeltaRJ'
\nonumber\\[2mm]
&=\yb+\Deltas\cdot[1+\Delta_{\rm r}]\cdot[\DeltaRJ\,(1+\Delta_{\rm r})+\Delta_{\rm r}]
\\%[2mm]
\label{eq:D_D2b'appr}
&\approx \yb+\yxxx +\Deltas\cdot\Delta_{\rm r}
\Abst{,}
\end{align}
\esub
%-----------
with $\Deltar=(\Tref-\Treft)/\Treft$ and where prime denotes, that the
reference temperature $\Treft$ was used. The difference of
y-parameters is given by $\Delta
y=\ydifs'-\ydifs=\Deltas\cdot\Deltar$. For $\Tref=\TRJ$ the
contribution $\yxxx=0$. Therfore we can write the inferred y-parameter
for any reference temperature as:
%-----------
\beal
\label{eq:y_inf_gen}
y(\Tref)\approx \yb+
\frac{\bar{T}_{\rm b,2}-\bar{T}_{\rm b,1}}{\bar{T}_{\rm b,1}+\bar{T}_{\rm b,2}}
\cdot\frac{\bar{T}_{\rm b,1}+\bar{T}_{\rm b,2}-2\,\Tref}{\Tref}
\Abst{.}
\end{align}
%------------
This result describes all the cases discussed previously: If we want
to discuss the beam distortions we have to set the reference
temperature to $\TRJ$ making $y\approx\yb$. If we are discussing the
y-parameter for the comparison of the SB with some reference blackbody
we have $\bar{T}_{\rm b,1}=\Tref$ resulting in
$y\approx\yb+\DeltaRJ^2/2$. This implies that in the limit of small
spectral distortions the y-parameter for the superposition of Planck
spectra is completely determined by the second beam moments and the
beam RJ temperatures.

%--------------------------------------------------------------------------------------
%--------------------------------------------------------------------------------------
%--------------------------------------------------------------------------------------
\section{Superposition of two Planck spectra} 
\label{sec:Sup2P}
%-------------------------------------------------------
In order to illustrate the main results obtained in the previous
Sect. we now discuss the spectral distortions arising due to the sum
and difference of two pure blackbodies with different temperatures
$T_1$ and $T_2$ and the corresponding intensities $I_1=\Ipl(T_1)$ and
$I_2=\Ipl(T_2)$ in the presence of a reference blackbody
$\Iref=\Ipl(\Tref)$ with temperature $\Tref$.
%-----------------------------------------------------
In principle one can directly apply the results obtained in
Sect. \ref{app:Theorie} and easily derive the equations describing
this situation, but in order to check the derivations of the previous
Sect. we here choose an approach starting from the expansion
\eqref{eq:DI_I_appr} taking only first order corrections into
account. Since the CMB temperature fluctuations are very small this
approximation is sufficient.

The results obtained for the sum of two blackbodies apply to the case
of an {\it absolute} measurement of the CMB sky, where the beam
contains only blackbodies with two equally weighted temperatures.
%---------------------------------
In the limit of narrow beams the results obtained for the difference
of two blackbodies can be directly used to discuss the effects arising
in {\it differential} measurements. In Sect. \ref{sec:Cali} we will
use some of the result of this Sect. in the discussion about cross
calibration issues.

\subsection{Sum of two Planck spectra} 
\label{app:Sum2P}
We want to describe the difference $\Delta \Is=\Is-\Iref$, as a
function of frequency. Here we defined the sum of the two blackbodies
by $\Is=(I_1+I_2)/2$. The easiest is to first calculate the difference
$\Delta I_i=I_i-\Iref$ since $\Delta\Is\equiv (\Delta I_1+\Delta
I_2)/2$. If we define $\Delta_i=(T_i-\Tref)/\Tref$ and $\x=h\nu/\kB
\Tref$ we can make use of equation \eqref{eq:DI_I_appr} and write the
relative difference of $\Is$ and $\Iref$ as
%-----------
\beal
\label{eq:Is_Iref}
%\frac{\Delta\Is}{\Iref}
\frac{\Is-\Iref}{\Iref}
%&=\frac{\Delta I_1+\Delta I_2}{2\,\Iref}\nonumber\\[1mm]
&\approx\frac{\x\,e^\x}{e^\x-1}\cdot\left[\frac{\Delta_1+\Delta_2}{2}
+ g_y(\x)\cdot\frac{y_1+y_2}{2}\right]
\Abst{.}
\end{align}
%-----------
Here we abbreviated the y-parameters as $y_i=\Delta^2_i/2$. Defining the RJ
temperature $\TRJ=(T_1+T_2)/2$ of the sum of the two blackbodies, the
sum of the first order terms and the y-parameters may be rewritten
like
%-----------
\bsub
\label{eq:Ds}
\beal
\label{eq:Ds1}
\Delta_{\rm RJ}&=\frac{\Delta_1+\Delta_2}{2}=\frac{\TRJ-\Tref}{\Tref}
\\
%\frac{y_1+y_2}{2}&\approx\frac{1}{8}\left[\frac{\Delta T^2}{\TRJ^2}+4\,\frac{(\TRJ-\Tref)^2}{\TRJ^2} \right]
\label{eq:Ds2}
\ysum&=\frac{y_1+y_2}{2}\approx\frac{1}{2}\left[\Delta_{\rm RJ}^2+\left(\frac{T_1-T_2}{T_1+T_2}\right)^2\right]
\Abst{.}
\end{align}
\esub
%-----------
From this it can be seen that if the reference temperature is set to
$\TRJ$ the temperature distortion corresponding to the first term in
\eqref{eq:Is_Iref} vanishes ($\DeltaRJ=0$) and only a minimal
y-distortion with y-parameter
%-----------
\beal
\label{eq:yb}
\yb=\frac{1}{2}\,\left[\frac{T_1-T_2}{T_1+T_2}\right]^2
%\Abst{.}
\end{align}
%-----------
is left. At low frequencies ($\x<\x_y)$ this reflects the fact that
the sum of RJ spectra is again a RJ spectrum.
%----------------------------------
Comparing \eqref{eq:Ds} to \eqref{eq:delref} one may conclude that
$\yb$ corresponds to the contribution of the second beam moment
$\left<\delta^2\right>$ to the y-parameter. Comparing \eqref{eq:yb} to
\eqref{eq:moments_twob} with $w_1=w_2=\frac{1}{2}$ confirms this
conclusion.

%---------------
\begin{figure}
\centering
%\resizebox{\hsize}{!}{\includegraphics[width=8.5cm]{./eps/yplot_1.eps}}
\includegraphics[width=\plotwd]
%\plotone
%{./eps/y_inf.eps}
%{./eps/y_inf_dipole.eps}
{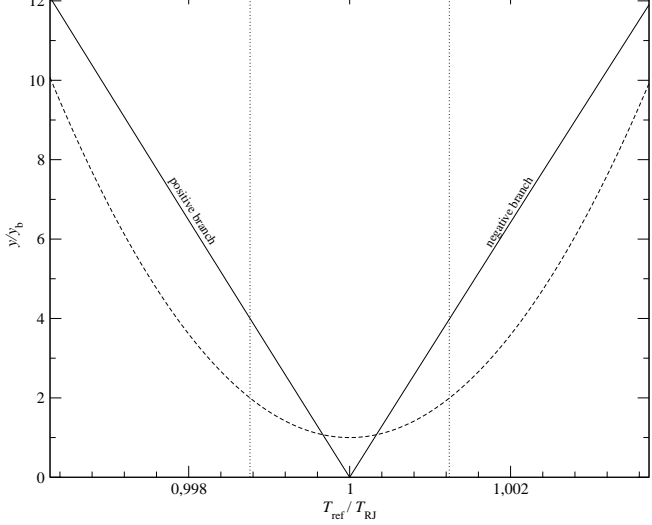}
\caption{Dependence of the inferred y-parameter on the chosen
reference temperature $\Tref$ for two blackbodies with temperatures
$T_1=2.72162\,$K and $T_2=2.72838\,$K, corresponding to the minimum
and maximum of the CMB dipole. In this case the mean RJ temperature is
$\TRJ=(T_1+T_2)/2=2.725\,$K and $\yb=7.7\cdot 10^{-7}$. --- Sum of two
blackbodies: total y-parameter as given by \eqref{eq:Ds2} (dashed
line). --- Difference of two blackbodies: the absolute value $|\ydif|$
of the inferred y-parameter as given in equation \eqref{eq:DD2} (solid
line). The dotted vertical lines show the location of $\Tref=T_1$ and
$\Tref=T_2$ respectively.}
\label{fig:y_inf}
\end{figure}
%------------------------
The dependence of the inferred y-parameter on the chosen reference
temperature is illustrated in Figure \ref{fig:y_inf}. As mentioned
above the y-parameter reaches a minimum for $\Tref=\TRJ$. It is
important to note that the minimal value of the y-parameter $\yb$ does
not depend on the reference temperature $\Tref$ but only on the second
beam moment of the temperature distribution function.
%-------------------------------------------
Changing $\Tref$ from $\TRJ \rightarrow T_i$ the inferred y-parameter
monotonically increases being a factor of $\sim 2$ larger than $\yb$
for $\Tref=T_i$. This can be understood as follows: If we set
$\Tref=T_i$ we obtain $\ysum=(T_1-T_2)^2/4\,T^2_i\approx
(T_1-T_2)^2/4\,T^2_1+\mathcal{O}(\Delta T^3/T_1^3)$, with $\Delta
T=T_1-T_2$. Therefore the ratio
$\ysum/\yd\approx(T_1+T_2)^2/2\,T^2_1\approx 2$.

%----------------------------------------------
Outside the region $T_1\leq\Tref\leq T_2$ the inferred y-parameter
increases further. It is in principle possible to gain large factors
in the inferred y-parameter by going very far away from the RJ
temperature of the SB, but at some point next order corrections will
become important and therefore limit $\Tref$ to the region where the
approximation of small spectral distortions is still valid. For the
CMB the most natural choices of the reference temperature are the
whole sky mean temperature $T_0$ and the maximum or minimum of the CMB
dipole. For the SB the inferred y-parameter varies from $y\sim
7.7\cdot 10^{-7}$ for $\Tref=T_0$ to $y\sim 1.5\cdot 10^{-6}$ for
$\Tref=T_1$. The behavior of the inferred y-parameter shows, that in
order to minimize the arising spectral distortion for {\it absolute}
measurements it is important to use an internal calibrator with
temperature close to the beam RJ temperature $\TRJ$.

%----------------------------------------------
In the limit $T_1\rightarrow T_2$ we are in principle comparing one
pure Planck spectrum with a reference blackbody. This case applies for
an {\it absolute} measurement of the CMB sky in the limit of very
narrow beams. In this case it follows that $\yb\rightarrow 0$ and in
addition $\ysum\rightarrow 0$ for $\Tref=\TRJ$. Except for some
changes in scales the behavior of the curve shown in
Fig. \ref{fig:y_inf} is unaffected.

\subsection{Difference of two Planck spectra} 
\label{app:Dif2P}
%------------
Since $\Delta I =I_2-I_1\equiv \Delta I_2-\Delta I_1$, where $\Delta
I_i=I_i-\Iref$, applying equation \eqref{eq:DI_I_appr} one can easily
find
%-----------
\beal
\label{eq:DI_2_Iref}
\frac{I_2-I_1}{\Iref}
%&=\frac{\Delta I_2-\Delta I_1}{\Iref}\nonumber\\[1mm]
&\approx\frac{\x\,e^\x}{e^\x-1}\cdot\Big[\Delta_2-\Delta_1
+ g_y(\x)\cdot(y_2-y_1)\Big]
\Abst{.}
\end{align}
%-----------
Again it is possible to simplify the difference of the first order
terms and the y-parameters leading to
 %-----------
\bsub
\label{eq:DD}
\beal
\label{eq:DD1}
\Delta_{12}&=\Delta_2-\Delta_1=\frac{T_2-T_1}{\Tref}
\\
%\frac{y_1+y_2}{2}&\approx\frac{1}{8}\left[\frac{\Delta T^2}{\TRJ^2}+4\,\frac{(\TRJ-\Tref)^2}{\TRJ^2} \right]
\label{eq:DD2}
\ydif&=y_2-y_1=\Delta_{12}\cdot\Delta_{\rm RJ}
\Abst{,}
\end{align}
\esub
%-----------
where $\Delta_{\rm RJ}=(\TRJ-\Tref)/\Tref$ and $\TRJ=(T_1+T_2)/2$. For
the difference of two Plancks the temperature distortion does not
vanish for any choice of the reference temperature (see equation
\eqref{eq:DD1}), whereas the y-distortion changes sign at
$\Tref=\TRJ$. 
%--------------------------
As mentioned earlier, this shows that for the difference of two Planck
spectra in order to obtain a positive y-parameter it is in general not
sufficient to choose the intensity with higher RJ temperature as
$I_2$. Comparing \eqref{eq:DD} to \eqref{eq:D_D2a} and
\eqref{eq:D2_D1} and keeping in mind that both beam moments
$\left<\delta_i^2\right>_{\rm b,\it i}=0$ shows the equivalence of
both approaches.
%----------------------------------------
In Figure \ref{fig:y_inf} the dependence of the absolute value of the
inferred y-parameter on the chosen reference temperature is shown. At
$\Tref=T_i$ the inferred y-parameter for the difference of two
blackbodies is 2 times larger than in the case of the sum,
$\left.(|\ydif|/\ysum)\right|_{T_i}\sim 2$.

As in the case of the sum of two Plancks for $\Tref$ outside the
region $T_1\leq\Tref\leq T_2$ the inferred y-parameter increases
strongly. Again there is a limit set by the approximations of small
spectral distortions. The behavior of the inferred y-parameter shows,
that in order to minimize the spectral distortion arising in
differential measurements it is important to choose a reference
temperature close to the RJ temperature of the observed region.

\section{Spectral distortions due to the CMB dipole} 
\label{app:dipole}
%------------
The largest temperature fluctuation on the observed CMB sky is
connected with the CMB dipole. Its amplitude has been accurately
measured by {\sc Cobe/Firas} \citep{Fixsen02}: $\Delta T_{\rm
d}=3.381\pm 0.007$mK, corresponding to $\Delta T/T\sim 0.2\,\%$ on
very large angular scales. Assuming that it is only due to the motion
of the solar system with respect to the CMB rest frame
%\footnote{According to theoretical predictions probably only $1\%$ is
%contributed by the intrinsic CMB dipole.} 
it implies a velocity of $v=372\pm 1$~km/s in the direction $(l,
b)=(264.14^\circ\pm 0.15^\circ, 48.26^\circ\pm 0.15^\circ)$. In order
to understand the spectral distortions arising due to the dipole, we
start with the direction dependent temperature of the CMB, neglecting
any intrinsic anisotropy:
%\citep{Peebles68}:
%-----------
\bsub
\beal
\label{eq:Tofd}
T(\mu)
&=\frac{T_0}{\gamma[1-\beta\,\mu]}
\\
\label{eq:Tsmallb}
&\!\!\stackrel{\stackrel{\beta\ll 1}{\downarrow}}{\approx} 
T_0\,\left[1-\frac{\beta^2}{6}+\beta\,\mu+\beta^2\left(\mu^2-\frac{1}{3}\right)\right]
\Abst{.}
\end{align}
\esub
%-----------
Here we introduced the abbreviation $\mu=\cos\theta$, with $\theta$
being the angle between the direction of $v$ and the location on the
sky. $T_0$ denotes the intrinsic CMB temperature, $\beta=v/c$ is the
velocity in units of the speed of light and
$\gamma=1/\sqrt{1-\beta^2}$ is the corresponding Lorentz factor. The
whole sky mean RJ temperature is given by
%-----------
\beal
\label{eq:Tmeanall}
\bar{T}_{\rm f}=\frac{T_0}{2\gamma\beta}\,\ln\left[\frac{1+\beta}{1-\beta}\right]
\stackrel{\stackrel{\beta\ll 1}{\downarrow}}{\approx} T_0\,\left[1-\frac{\beta^2}{6}\right]
\Abst{,}
\end{align}
%-----------
where 'f' denotes full sky average temperature. Equation
\eqref{eq:Tmeanall} implies that due to the motion of the solar system
relative to the CMB restframe the observed whole sky mean RJ
temperature is $\Delta T=0.70\,\mu$K less than the intrinsic
temperature $T_0$. The measured value of the whole sky mean %RJ
temperature as given by the {\sc Cobe/Firas} experiment
\citep{Fixsen02} is $T_0=2.725\pm 0.001\,$K. Therefore this difference
is still $\sim 1000$ times smaller than the current error bars and far
from being measured.

Below we now discuss the spectral distortions arising due to the CMB
dipole in the context of finite angular resolution. Since only the
second moment of the temperature distribution is important, we will
use the second order expansion of $T(\mu)$ in $\beta$ as given in
equation \eqref{eq:Tsmallb}.
%-----------------------------------
In the last part of this section we show that, starting with equation
\eqref{eq:Tofd} all the results of this section can also be directly
obtained by expansion of the blackbody $\Ipl(T(\mu))$ in terms of
small $\beta$. This shows the equivalence of both
approaches. Nevertheless, the big advantage of the treatment developed
in Sect. \ref{app:Theorie} is that it can be applied to general
temperature distributions and that the source of the y-distortion can
be directly related to the second moment of the temperature
distribution.

%Below we discuss the spectral distortion arising due to the CMB dipole
%in the context of finite angular resolution. Since only the second
%moment of the temperature distribution will be important, we expand
%equation \eqref{eq:Tofd} in terms of $\beta$ up to second order \citep{Sun80}:
%%-----------
%\beal
%\label{eq:Tsmallb}
%T(\mu)=T_0\,\left[1-\frac{\beta^2}{6}+\beta\,\mu+\beta^2\left(\mu^2-\frac{1}{3}\right)\right]
%\Abst{.}
%\end{align}
%%-----------
%Furthermore, we show that, using equation \eqref{eq:Tofd} all the
%results can also be directly obtained by expansion of the blackbody
%$\Ipl(T(\mu))$ in terms of small $\beta$. This shows the equivalence
%of both approaches. Nevertheless, a big advantage of the treatment
%developed above is, that it can be applied to general temperature
%distributions and that the source of any spectral distortion can be
%directly related to the second moment of the temperature distribution.

\subsection{Whole sky beam spectral distortion}
Defining $\delta=[T(\mu)-\bar{T}_{\rm f}]/\bar{T}_{\rm f}$, where
$\bar{T}_{\rm f}$ is given by equation \eqref{eq:Tmeanall}, the full
sky moments can be calculated by the integrals
%-----------
$\left<\delta^k\right>_{\rm f}=\frac{1}{2}\int^1_{-1}\delta(\mu)^k\id \mu$. 
%-----------
The first three moments may be found as
%-----------
\bsub
\beal
\label{eq:del2}
\left<\delta^0\right>_{\rm f}&=1, \qquad \left<\delta^1\right>_{\rm f}=0
\\
\left<\delta^2\right>_{\rm f}&=\frac{4\beta^2\gamma^2}{\Xi^{2}}-1
\stackrel{\stackrel{\beta\ll 1}{\downarrow}}{\approx}
\frac{\beta^2}{3}+\mathcal{O}(\beta^4)
%+\frac{4\,\beta^4}{15} +\mathcal{O}(\beta^6)
%\\
%\left<\delta^3\right>_{\rm f}&=\frac{8\beta^3\gamma^4}{\Xi^{3}}-\frac{12\beta^2\gamma^2}{\Xi^{2}}+2
%\stackrel{\stackrel{\beta\ll 1}{\downarrow}}{\approx}
%\frac{4\,\beta^4}{15}+\mathcal{O}(\beta^6)
%\\
%\left<\delta^4\right>_{\rm f}&=\frac{16\beta^4[3+\beta^2]\gamma^6}{3\,\Xi^{4}}
%-\frac{32\beta^3\gamma^4}{\Xi^{3}}
%+\frac{24\beta^2\gamma^2}{\Xi^{2}}
%-3
%\nonumber\\
%&\!\!\stackrel{\stackrel{\beta\ll 1}{\downarrow}}{\approx}
%\frac{\beta^4}{5}+\mathcal{O}(\beta^6)
\Abst{,}
\end{align}
\esub
%-----------
where $\Xi=\ln\!\!\big[\frac{1+\beta}{1-\beta}\big]$. With $\beta_{\rm
d}=\Delta T_{\rm d}/T_0=1.241 \cdot 10^{-3}$ one finds:
$\left<\delta^2\right>_{\rm f}=5.1\cdot10^{-7}$. This implies a full
sky y-distortion with
%-----------
%$\yd=\beta_{\rm d}^2/6 = 2.6\cdot10^{-7}$.
\beal
\label{eq:y_d}
\yd=\frac{\beta_{\rm d}^2}{6}\approx 2.6\cdot10^{-7}
\Abst{,}
\end{align}
%-----------
which is currently $\sim 60$ times below the {\sc Cobe/Firas} upper
limit. This distortion translates into a temperature difference of
$\Delta T=0.7\cdot g_y(\x)\,\mu$K. The exact behavior of $g_y(\x)$ is
shown in Fig. \ref{fig:four}.
%-------------------------------------------
%The corresponding absolute temperature difference in some of the {\sc
%Planck} frequency channels are given in Table \ref{tab:one}. 
%-------------------------------------------
We have checked that deviations from a y-distortion will become
important only at extremely high frequencies ($\x\geq 360$).

Here we want to note, that this full sky distortions arises due to the
existence of the dipole anisotropy. In absolute measurements of the
CMB $\yd$ places a lower limit on the full sky y-parameter (see
Sect. \ref{sec:SumP}).

\subsection{Beam spectral distortion due to the CMB dipole}
Here we are interested in the angular pattern of the y-distortion
induced by the SB over the dipole for an observation with a finite
angular resolution and in particular in the location of the maximal
y-distortion, when we compare the beam flux at different frequencies
to a reference blackbody with beam RJ temperature. To model the beam
we use a simple top-hat filter function
%-----------
\beal
\label{eq:Tophat}
W(\theta')=
\begin{cases}
1 &\text{for $\theta'\leq \thr$,}\\
0 &\text{else,}
\end{cases}
\end{align}
%-----------
where the $z'$-axis is along the beam direction\footnote{Prime denotes
coordinates with respect to the system $S'$ where the $z'$-axis is
defined by the direction of the beam.} and $\thr$ is the radius of the
top-hat in spherical coordinates.

The $k$th moment of some variable $X$ over the beam is then given by
the integral
%-----------
\beal
\label{eq:Mom}
\left<X^k\right>_{\rm r}=\frac{1}{2\,\pi\,(1-\mu_{\rm r})}\int^{2\,\pi}_0\int^1_{\mu_{\rm r}}X^k
\id \mu'\id \phi'
\Abst{,}
\end{align}
%-----------
where 'r' denotes beam average for a top-hat of radius $\thr$ and
$\mu_{\rm r}=\cos\thr$. Defining $\mu_0=\cos\theta_0$ and using
%the relation
%-----------
\beal
\label{eq:mu_tpp}
\mu=\mu_0\cdot\mu'+\cos(\phi'-\phi_0)\,\sin(\theta_0)\sqrt{1-\mu'^2}
\Abst{,}
\end{align}
%-----------
one can calculate the beam averages of the dipole and quadrupole
anisotropy
%-----------
\bsub
\label{eq:mu_mom}
\beal
\left<\mu\right>_{\rm r}
&=\zeta_+\,\mu_0\\
\left<\mu^2-\frac{1}{3}\right>_{\rm r}
&=\mu_{\rm r}\,\zeta_+\!\left(\mu_0^2-\frac{1}{3}\right)
\Abst{,}
\end{align}
\esub
%-----------
where we introduced the abbreviation $\zeta_\pm=\frac{1\pm\mu_{\rm
r}}{2}$. These equation will be very useful in all the following
discussion.

%-------------------------
Using formula \eqref{eq:Tsmallb} and equations \eqref{eq:Mom} and
\eqref{eq:mu_mom} the mean RJ temperature inside the beam in some
direction $(\phi_0,\theta_0)$ relative to the dipole axis is given as
%-----------
\beal
\label{eq:T_m_d}
\bar{T}_{\rm r}=T_0\,\left[1-\frac{\beta^2}{6}+\beta\,\zeta_+\,\mu_0
+\,\beta^2\,\mu_{\rm r}\,\zeta_+\!\left(\mu_0^2-\frac{1}{3}\right)\right]
\Abst{.}
\end{align}
%-----------
The position of the maximum and minimum is, as expected, at
$\theta_0=0$ and $\theta_0=\pi$ respectively. 
%, i.e. along and in the opposite direction of the dipole moment.
%-----------

%---------------
\begin{figure}
\centering
%\resizebox{\hsize}{!}{\includegraphics[width=8.5cm]{./eps/yplot_1.eps}}
\includegraphics[width=8.8cm]
%\plotone
%{./eps/my1b.ps}
{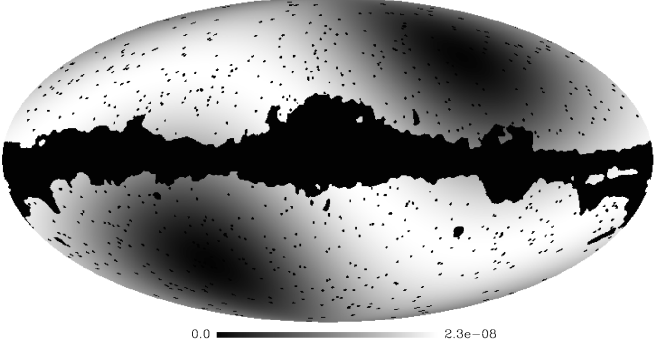}
\caption{Angular distribution of $\yc$ as given by equation
\eqref{eq:yc} for $\thr=20^\circ$. The maximal spectral distortion is
expected to appear in a ring perpendicular to the dipole axis and in
this case has a value $y_{\rm r, max}=2.3\cdot 10^{-8}$. The minimal
y-parameter has a value of $y=2.4\cdot 10^{-10}$ and is located around
the maximum and minimum of the dipole. The galactic plane was cut out
using the kp0-mask of the {\sc Wmap} data base.}
\label{fig:mone}
\end{figure}
%---------------
Now the second moment of $\delta=[T(\mu)-\bar{T}_{\rm r}]/\bar{T}_{\rm
r}$ inside the beam can be derived as:
%-----------
\beal
\label{eq:del2_d}
\left<\delta^2\right>_{\rm r}
&\!\!\!\stackrel{\stackrel{\beta_{\rm d}\ll 1}{\downarrow}}{\approx}
\frac{\beta^2_{\rm d}}{3}\cdot\left[1+\zeta_+
-3\,\zeta_+\!\left(\mu_0^2-\frac{1}{3}\right)\right] \zeta_-
\Abst{,}
\end{align}
%-----------
where up to second order in $\beta$ only the CMB dipole contributes
and we have set $\beta\equiv \beta_{\rm d}$. The corresponding
y-parameter 
%-----------
\beal
\label{eq:yc}
\yc=\yd\cdot\left[1+\zeta_+ -3\,\zeta_+\!\left(\mu_0^2-\frac{1}{3}\right)\right] \zeta_-
\Abst{,}
\end{align}
%-----------
has a monopole and quadrupole angular dependence. As expected it
vanishes for high angular resolution ($\zeta_-\rightarrow 0$). In
Fig. \ref{fig:mone} the angular distribution of $\yc$ for
$\thr=20^\circ$ is illustrated. The maximum lies in a broad ring
perpendicular to the dipole axis, which even without taking the galaxy
into account is covering a large fraction of the sky. In the shown
case, the maximal value of the y-parameter is approximately $y_{\rm r,
max}\sim\yd/10$.

\begin{figure}
%\centering
%\resizebox{\hsize}{!}{\includegraphics[width=8.5cm]{./eps/yplot_1.eps}}
%\includegraphics[width=\plotwd]
\includegraphics[width=8.5cm]
%\plotone
%{./eps/y0_strat.eps}
{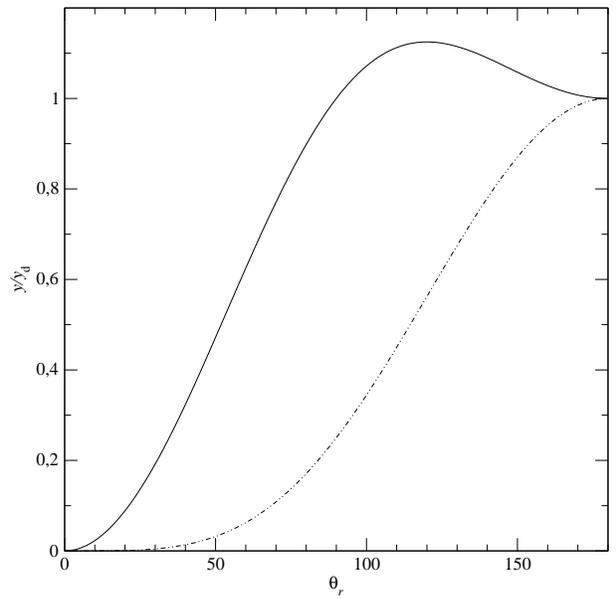}
\caption{Dependence of the y-parameter in units of $\yd=2.6\cdot
10^{-7}$ on the beam radius $\thr$ in degree for a circular beam: $\yc$
for $\mu_0=\pm1$ as given by equation \eqref{eq:yc} (dash-dot-dotted),
$y_{\rm r, max}$ according to equation \eqref{eq:y_max_r} (solid). For
values $-1<\mu_0<1$ the curves lie between these two extremes.}
\label{fig:zero}
\end{figure}

Using equation \eqref{eq:del2_d} the position of the maximum can be
found: $\theta_{0, \rm max}=\frac{\pi}{2}$.  This suggests that the
location of the maximal distortion is where the derivative of the
temperature distribution is extremal. The maximal y-parameter is given
as
%-----------
\beal
\label{eq:y_max_r}
%y_{\rm r, max}= {\yd}\cdot\left[1-\frac{\mu_{\rm r}(\mu_{\rm r}+1)}{2}\right]
y_{\rm r, max}= {\yd}\cdot\left[1-\mu_{\rm r}\,\zeta_+\right]
\Abst{.}
\end{align}
%-----------
The dependence of $y_{\rm r, max}$ on the beam radius is shown in Fig.
\ref{fig:zero}.
%---------------
%------------------------
%In general, the distortion inside a single beam is less than the full sky
%distortion, $y_{\rm r}<{\yd}$. But as mentioned earlier, upon averaging
%over the whole sky the total distortion again will be given by ${\yd}$.
%---------------
For beam radii larger than $\thr=90^\circ$ it follows that $y_{\rm r,
max}\geq {\yd}$ and at $\thr=120^\circ$ it is $9/8$ times ${\yd}$. The
maximal y-parameter, $y_{\rm r, max}$, is a steep function of beam
radius: It has values $y_{\rm d}/2$, ${\yd}/3$ and ${\yd}/10$ for
$\thr=51.83^\circ$, $\thr=40.68^\circ$ and $\thr=21.29^\circ$
respectively and vanishes like $\sim 3\,\thr^2/4$ for high angular
resolution. For $\thr=5^\circ$ it is $\sim 6\cdot 10^{-3}\times
{\yd}$. This implies that the beam spectral distortions resulting from
the dipole are lower than $\Delta T/T\sim 10^{-9}$ for experiments
with resolution better than $\sim 10^\circ$. As will be shown later,
on this level contributions from the higher multipoles will start to
become important (Sect. \ref{sec:yWMAP}). 

\subsection{Distortion with respect to any $\Tref$}
\label{sec:BeamTref}
%------------------------------------
We have seen in Sect. \ref{sec:SumP} that the inferred y-parameter for
the comparison of the SB with some reference blackbody with
temperature $\Tref$ has one contribution from the second beam moment
$\left<\delta^2\right>$ and another due to the dispersion $\DeltaRJ^2$
of the beam RJ temperature relative to the temperature of the chosen
reference blackbody (cf. equation \eqref{eq:delrefb}). Therefore one
can immediately write down the relative difference of the beam
intensity and the intensity of the chosen reference blackbody making
use of \eqref{eq:delref}, \eqref{eq:nt_appr_ref}, \eqref{eq:T_m_d} and
\eqref{eq:del2_d}. In the limit $\thr\rightarrow 0$
we can apply the results obtained in Sect. \ref{app:Sum2P} for the sum
of two blackbodies, if we set $T_1=T_2$.
%-------------------------------------------
Here we want to discuss the case $\Tref=T_0$ in more detail. For this
choice the residual full sky y-distortion (which may be obtained by
averaging the intensity data over the whole sky) is minimal.
%-------------------------------------------

\subsection*{Distortion with respect to $T_0$}
Now, instead of comparing the spectrum of the SB to a blackbody with
beam RJ temperature $\bar{T}_{\rm r}$ we compare to a blackbody of
temperature $T_0$. Then the spectral distortion follows from equation
\eqref{eq:nt_appr_ref} and in addition to the beam moment
\eqref{eq:del2_d} the first two moments of $\Delta=(\bar{T}_{\rm
r}-T_0)/T_0$ have to be calculated:
%-----------
\bsub
\label{eq:D}
\beal
\label{eq:D_a}
\Delta^1
&= -\frac{\beta^2_{\rm d}}{6}+\beta_{\rm d}\,\zeta_+\,\mu_0
+\,\beta_{\rm d}^2\,\mu_{\rm r}\,\zeta_+\!\left(\mu_0^2-\frac{1}{3}\right)
\\
\label{eq:D_b}
\Delta^2
&=\;\;\,\frac{\beta^2_{\rm d}}{3}\,\zeta_+^2
+\beta^2_{\rm d}\,\zeta_+^2\!\left(\mu_0^2-\frac{1}{3}\right)
\Abst{.}
\end{align}
\esub
%-----------
The spectral distortion arising from the SB can now be characterized as
follows:
%-----------
%Term $\propto \Delta^1$: 
(i) At low frequencies the y-distortion vanishes. The motion of the
solar system with respect to the CMB restframe induces a temperature
dipole ($\sim\beta_{\rm d}$) and a temperature monopole and quadrupole
($\sim\beta^2_{\rm d}$) all resulting from the $\Delta^1$-term:
%-----------
\beal
\label{eq:DT_the}
\left.\frac{\Delta T}{T_0}\right|_{\rm t}
=-\frac{\beta^2_{\rm d}}{6}+\beta_{\rm d}\,\zeta_+\,\mu_0
+\,\beta_{\rm d}^2\,\mu_{\rm r}\,\zeta_+\!\left(\mu_0^2-\frac{1}{3}\right)
\Abst{.}
\end{align}
%-----------
In the limit of high angular resolution ($\zeta_+, \mu_{\rm
r}\rightarrow 1$) this is equivalent to the expansion
\eqref{eq:Tsmallb}. (ii) A y-distortion is induced which is
proportional to the sum $\Delta^2+\left<\delta^2\right>_{\rm r}$:
%-----------
\beal
\label{eq:DT_x}
\left.\frac{\Delta T}{T_0}\right|_{\rm y}
=\frac{g_2(\x)}{g_1(\x)}\,\left[
\frac{\beta^2_{\rm d}}{3}+\,\beta_{\rm d}^2\,\mu_{\rm r}\,\zeta_+\!\left(\mu_0^2-\frac{1}{3}\right)
\right]
\Abst{.}
\end{align}
%-----------
It has a monopole and quadrupole angular dependence and only arises
due to the CMB dipole. The sum of both contributions mentioned above
can be rewritten as
%-----------
\bsub
\label{eq:DT}
\beal
\left.\frac{\Delta T}{T_0}\right|_{\rm tot}
%&=\left.\frac{\Delta T}{T_0}\right|_{\rm th}+\frac{\Delta T(\x)}{T_0}
%\nonumber\\
\label{eq:DT_a}
&=\Delta-\Delta^2-\left<\delta^2\right>_{\rm r}+\frac{g(\x)}{2}\,\left[\Delta^2+\left<\delta^2\right>_{\rm r}\right]
\\
&=-\frac{\beta^2_{\rm d}}{2}+\beta_{\rm d}\,\zeta_+\,\mu_0
\nonumber\\
&\qquad+\frac{g(\x)}{2}\,\left[
\frac{\beta^2_{\rm d}}{3}+\,\beta_{\rm d}^2\,\mu_{\rm r}\,\zeta_+\!\left(\mu_0^2-\frac{1}{3}\right)
\right]
\Abst{,}
\end{align}
\esub
%-----------
where we used the definition \eqref{eq:g} for $g(\x)$.
%where we have defined the function
%%-----------
%\beal
%\label{eq:g}
%g(\x)=\frac{\x}{2}\,\frac{e^{\x}+1}{e^{\x}-1}
%\Abst{.}
%\end{align}
%%-----------
Rewriting the function $g_2(\x)$ in terms of $g(\x)$ lead to exact cancellation
of the motion induced temperature quadrupole and changes the temperature
monopole by $\bd^2/3$ (cf. $\Delta-\Delta^2-\left<\delta^2\right>_{\rm r}$-term in equation
\eqref{eq:DT_a}).

Here we want to note, that in the picture of the SB it is easily
understandable, that there is no difference in the resulting spectral
distortion whether the dipole anisotropy is intrinsic or due to
motion. Therefore, as has been noted earlier by \citet{Kam03}, it is
impossible to distinguish the intrinsic dipole from a motion induced
dipole by measurement of the frequency dependent temperature
quadrupole. 

\subsection*{Expansion of $\Delta I/I_0$ for small $\beta$}
\label{app:equiv}
%------------------------------------
Here we want to show, that the results \eqref{eq:del2_d} and
\eqref{eq:DT} can also be directly obtained starting with the
expansion of the blackbody spectrum with temperature $T(\mu)$ for
small velocity $\beta$ and thereby prove the equivalence and
correctness of both approaches. 
%---------
For this, inserting \eqref{eq:Tofd} into the blackbody spectrum
\eqref{eq:I_pl}, by Taylor expansion up to second order of $\beta$ one
may find
%-----------
\bmulti
\label{eq:DI_I0_tot}
\frac{\Delta I}{I_0}
=\beta^2\,g_1(\x)\,\left[\frac{g(\x)}{3}-\frac{1}{2}\right]+\,\beta\,g_1(\x)\,\mu
\\
+\beta^2\,g_1(\x)\,g(\x)\!\left(\!\mu^2-\frac{1}{3}\right)
\Abst{.}
\end{multline}
%-----------
This equation has been obtained earlier by
%\cite{Sun81} and by 
\citet{Saz1999} for a cluster of galaxies moving with respect to the
CMB and was later applied by \citet{Kam03} to discuss aspects of the
observed CMB dipole and quadrupole. It is valid for a measurement of
the CMB temperature with high angular resolution, where the
y-parameter due to the second beam moment is negligible ($y\sim
10^{-9}$ for $10^\circ$ angular resolution).

First we want to calculate the SB in a circular beam and compare it to
the blackbody $I_0$ of temperature $T_0$, i.e. we want to derive
$\left<\Delta I/I_0\right>_{\rm r}$. Using equation \eqref{eq:Mom} and
since only $\mu$ and $\mu^2$ depend on $\mu'$ and $\phi'$ this
integration with \eqref{eq:mu_mom} immediately leads to equation
\eqref{eq:DT}.

Next we want to derive the spectral distortion relative to the
reference blackbody with beam RJ temperature starting with
\eqref{eq:DI_I0_tot}. For this we rearrange \eqref{eq:DI_I0_tot} in
terms of $g_1(\x)$ and $g_2(\x)$ leading to
%-----------
\beal
\label{eq:DI_I0_tot_g1_2}
I(\x)=I_0(\x)\cdot[1+g_1(\x)\cdot \delta_0 + g_2(\x)\cdot \delta_0^2 ]
\Abst{,}
\end{align}
%-----------
with
$\delta_0=-\frac{\beta^2}{6}+\beta\,\mu+\beta^2\left(\mu^2-\frac{1}{3}\right)$.
%-------------------------------
In the RJ limit $g_1\rightarrow 1$ and $g_2\rightarrow 0$. This
immediately leads from equation \eqref{eq:DI_I0_tot_g1_2} to $T_{\rm
RJ}=\left<T_0\cdot[1+\delta_0]\right>_{\rm r}=T_0\cdot[1+\Delta]$,
where $\Delta$ is given by equation \eqref{eq:D_a}.
%-----------
Now we replace $\x=h\,\nu/k\,T_0=(h\,\nu/k\,\TRJ)\cdot T_{\rm
RJ}/T_0=\x_\ast\cdot[1+\Delta]$ in equation \eqref{eq:DI_I0_tot_g1_2}
and expand in second order of $\Delta$ obtaining:
%-----------
\bmulti
\label{eq:DI_Istar}
I(\x_\ast)=I_\ast(\x_\ast)\cdot\left[1+g_1(\x_\ast)\cdot(\delta_0-\Delta+\Delta^2-\delta_0\Delta)
\right.
\\
\left.
+\,g_2(\x_\ast)\cdot(\delta_0^2+\Delta^2-2\,\delta_0\Delta)\right]
\Abst{,}
\end{multline}
%-----------
with $I_\ast(\x_\ast)=\Ipl(\TRJ)$. Since only $\delta_0$ depends
on $\mu'$ and $\phi'$ we may perform the beam averages using equations
\eqref{eq:Mom} and \eqref{eq:mu_mom}. This is equivalent to replacing
$\delta_0\rightarrow \Delta$ and $\delta_0^2\rightarrow
\left<\delta_0^2\right>_{\rm r}$ in equation
\eqref{eq:DI_Istar}. Therefore this leads to
%-----------
\beal
\label{eq:DI_Istar_res}
\frac{I-I_\ast}{I_\ast}=g_2(\x_\ast)\cdot[\left<\delta_0^2\right>_{\rm r}-\Delta^2]
\Abst{.}
\end{align}
%-----------
Inserting equation \eqref{eq:D_b} and
%-----------
\beal
\label{eq:del0_2}
\left<\delta_0^2\right>_{\rm r}
=\frac{\beta^2}{3} +\beta^2\,\mu_{\rm r}\,\zeta_+\!\left(\mu_0^2-\frac{1}{3}\right)
\end{align}
%-----------
one may obtain $\left<\delta_0^2\right>_{\rm
r}-\Delta^2\equiv\left<\delta^2\right>_{\rm r}$, where
$\left<\delta^2\right>_{\rm r}$ is given by equation \eqref{eq:del2_d}
and we made use of the identities
%-----------
\bsub
\label{eq:Id}
\beal
1-\zeta_+^2&=[1+\zeta_+]\,\zeta_-
\\
\mu_{\rm r}-\zeta_+ &=-\zeta_-
\Abst{.}
\end{align}
\esub
%-----------
Let us note here, that since only $\delta_0$ depends on $\mu'$ and
$\phi'$ we may have calculated the beam average already from equation
\eqref{eq:DI_I0_tot_g1_2}. After expansing in terms of $\Delta$ this
would have directly lead to the result \eqref{eq:DI_Istar_res} without
the intermediate step \eqref{eq:DI_Istar}. This finally shows the
complete equivalence of both approaches.

\section{Spectral distortions due to higher multipoles}
\label{sec:yWMAP}
%-----------
The largest temperature anisotropy on the CMB sky is connected with
the CMB dipole which was discussed in detail above
(Sect. \ref{app:dipole}). In what follows here we are only concerned
with spectral distortions arising from multipoles with $l\geq 2$. As
has been shown above, the spectral distortions induced by the CMB
dipole are indistinguishable from a y-distortion. Since the typical
amplitude of the temperature fluctuations for $l\geq 2$ is a factor of
$\sim 100$ less than the dipole anisotropy ($\Delta T/T\sim 10^{-5}$),
the spectral distortion arising from higher multipoles will also be
indistinguishable from a y-distortion.

\subsection*{Calculations using a realization of the CMB sky}
In order to investigate the spectral distortions arising from the
multipoles $l\geq 2$, we use one realization of the CMB sky computed
with the {\sc Synfast} code of the {\sc Healpix}
distribution\footnote{http://www.eso.org/science/healpix/} given the
theoretical $C_l$'s for the temperature anisotropies computed with the
{\sc Cmbfast} code\footnote{http://www.cmbfast.org/} for the WMAP best
fit model \citep{WMAP_params}. Here we are only interested in the
distortions arising from the primordial CMB anisotropies. Future
experiments will have $\sim\,$arcmin resolution. Therefore we perform
our analysis in the domain, $l\leq 3000$, corresponding to angular
scales with dimensions $\theta\gtrsim 3.6'$.

%-----------
Using the generated CMB maps we extract the temperature distribution
function $R(T)$ inside a circular beam of given radius $\thr$ in
different directions on the sky. We then calculate the spectrum and
the spectral distortion inside the beam using equation \eqref{eq:nt}
and setting the reference temperature to the beam RJ temperature
$\TRJ=\int R\,T \id T$. As expected, we found that the spectral
distortions are indistinguishable from a y-distortion with y-parameter
given by equation \eqref{eq:y_Sup}. As discussed above this follows
from the fact, that the temperature fluctuations for $l\geq 2$ are
extremely small ($\sim 10^{-5}$).

%---------------
\begin{figure}
\centering
%\resizebox{\hsize}{!}{\includegraphics[width=8.5cm]{./eps/yplot_1.eps}}
\includegraphics[width=\plotwd]
%\plotone
{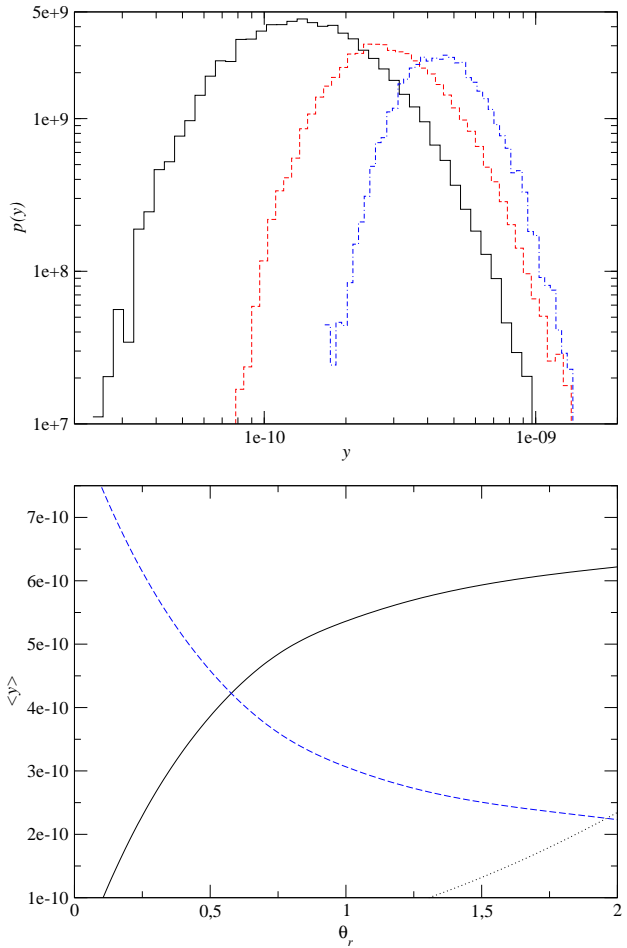}
\caption{Top: Probability density $p(y)$ for different aperture radii:
$\thr=0.25^\circ$ (solid), $\thr=0.5^\circ$ (dashed) and
$\thr=1.0^\circ$ (dashed-dotted). --- Bottom: Dependence of the
average y-parameter on the aperture radius $\thr$ in degrees: for the
{\it beam} spectral distortion (solid) and for the y-parameter
resulting from the dispersion of the beam RJ temperatures in different
directions on the sky relative to $T_0$ (dashed). Also shown is the
maximal beam spectral distortion arising due to the dipole, $y_{\rm r,
max}$ (dotted), according to equation \eqref{eq:y_max_r}.}
\label{fig:one}
\end{figure}
%---------------
In Fig. \ref{fig:one} the probability density $p(y)$ for a given
aperture is shown. It is defined such that $p(y)\,dy$ gives the
probability to find a value of the y-parameter in a random direction
on the sky between $y$ and $y+dy$ for a given beam radius.
%It is close to a $\chi^2$-distribution. 
As expected, the average value $\left<y\right>$ increases with
aperture radius, while the width of $p(y)$ decreases.
%This reflects the steep drop of power on small angular scales.  
For $\thr\gtrsim 1$ the average beam y-parameter varies only
slowly. In the limit of a whole sky measurement, the y-parameter
converges to $y_{\rm h}=8.33\cdot 10^{-10}$, which corresponds to the
whole sky rms dispersion $\Delta T\sim 111\,\mu$K of the used
realization of the CMB sky. For $\thr\lesssim 1$ the beam y-parameter
falls off very fast. As Fig. \ref{fig:one} suggests, for $\thr\lesssim
2- 5^\circ$ the distortions arising due to the dipole are negligible
in comparison to the distortions arising due to higher multipoles.

Figure \ref{fig:one} also shows the y-parameter resulting from the
dispersion of the beam RJ temperatures in different directions on the
sky relative to $T_0$. It drops with increasing beam radius, because
the average beam RJ temperature becomes closer to $T_0$ for bigger
beam radius. Looking at the sum of the {\it beam} distortion and the
distortion due to the dispersion of the beam RJ temperatures shows
that the averaged whole sky distortion is independent of the angular
resolution. 

%---------------
\begin{figure}
\centering
%\resizebox{\hsize}{!}{\includegraphics[width=8.5cm]{./eps/yplot_2.eps}}
\includegraphics[width=\plotwd]
%\plotone
%{./eps/yplot_2.eps}
{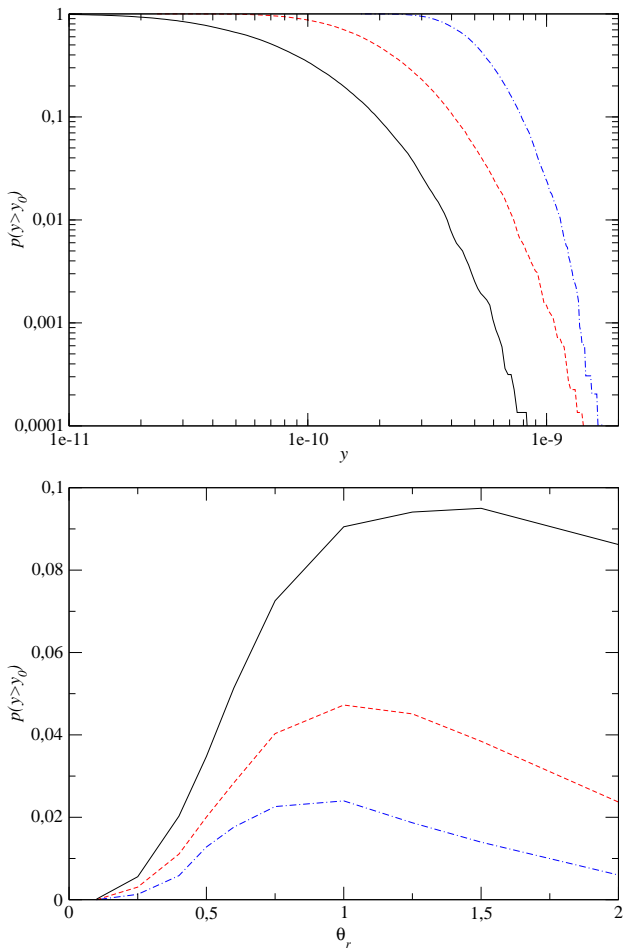}
%{./eps/y2.ps}
\caption{Top: Cumulative probability $p(y\geq y_0)$ for different aperture
radii
%---------------
$\thr=0.1^\circ$ (solid), $\thr=0.25^\circ$ (dashed) and $\thr=1.0^\circ$
(dashed-dotted). -- Bottom: Dependence of the cumulative probability
$p(y\geq y_0)$ on the aperture $\thr$ for $y_0=8\cdot 10^{-10}$ (solid),
$y_0=9\cdot 10^{-10}$ (dashed) and $y_0=10^{-9}$ (dashed-dotted).}
\label{fig:two}
\end{figure}
\begin{figure}
\centering
%\resizebox{\hsize}{!}{\includegraphics[width=8.5cm]{./eps/yplot_2.eps}}
\includegraphics[width=\plotwd]
%\plotone
{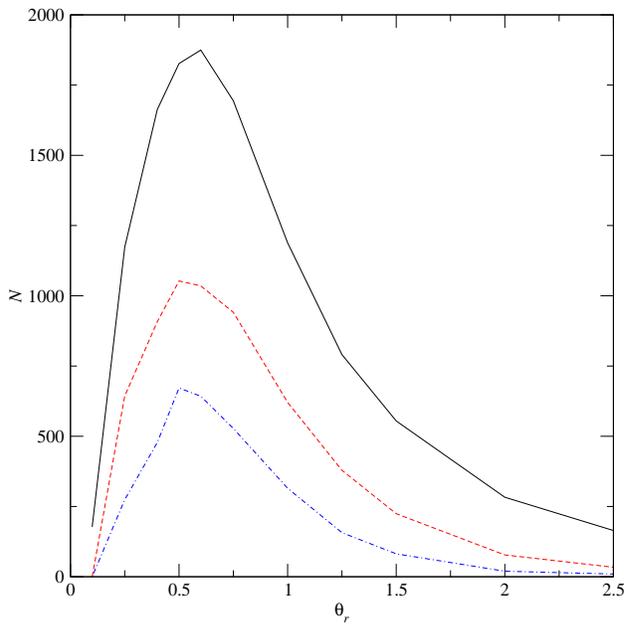}
\caption{Number of regions on the sky with $y\geq y_0$ for different
aperture radii $\thr$ in degree:
%---------------
for $y_0=8\cdot 10^{-10}$ (solid), $y_0=9\cdot 10^{-10}$ (dashed) and
$y_0=10^{-9}$ (dashed-dotted).}
\label{fig:three}
\end{figure}

%---------------
In Fig. \ref{fig:two} the cumulative probability for finding a
spectral distortion with $y\geq y_0$ is shown for different
apertures. Fixing the value $y_0$, there is a maximum probability of
$2.4\%$ to measure $y\geq 10^{-9}$ for an aperture radius $\thr\sim
1^\circ$ during the mapping or scanning of extended regions of the
sky. This correspond to $\sim 500$ sources on the whole sky. In Fig.
\ref{fig:three} the dependence of the number of sources with $y$ above
$y_0$ for a given radius $\thr$ is illustrated. It peaks around
$\thr\sim 0.5$ degree, corresponding to the first acoustic peak. 
%---------------
We should mention that the maxima of the y-distortion do not coincide
with the maxima of $\Delta T/T$, but that they both should have
similar statistical properties. We expect that the maxima of the
y-parameter are there, where the derivatives of the averaged
temperature field are large. 
%---------------
%---------------

\subsection*{The motion induced CMB octopole}
In section \ref{app:dipole} we discussed spectral distortions due to
the motion of the solar system relative to the CMB restframe up to
second order in $\bd$. On the level of $y\sim 10^{-10}-10^{-9}$ third
order corrections $\propto \bd^3$ should also start to contribute. In
this order of $\bd$ not only the motion induced dipole and quadrupole
lead to spectral distortions but also terms connected to the products
of the intrinsic dipole and quadrupole with the motion induced dipole
and quadrupole. Here we now only discuss the motion induced CMB third
order terms.

Expanding \eqref{eq:Tofd} up to third order in $\bd$ and calculating
the beam averaged RJ temperature for the top-hat beam using
\eqref{eq:Mom} leads to
%-----------
\bmulti
\label{eq:T_m_oct}
\bar{T}_{\rm r, oct}=\bar{T}_{\rm r}+T_0\,\bd^3\,\zeta_+\,\left[\frac{1}{10}\,\mu_0
\right.
\\
\left.
+\left(1-5\,\zeta_+\,\zeta_- \right)\mu_0\!\left(\mu_0^2-\frac{3}{5}\right)\right]
\Abst{,}
\end{multline}
%-----------
where $\bar{T}_{\rm r}$ is given by \eqref{eq:T_m_d}. In addition to
the motion induced octopole in third order of $\bd$ there also is a
correction to the CMB dipole.
%-----------
Now we can also recalculate the second beam moment leading to a
y-parameter of
%-----------
\bmulti
\label{eq:yc_oct}
y_{\rm r, oct}=\yd\,\bd \cdot
\left[\mu_0\left(1+\frac{7}{5}\,\mu_{\rm r}\right)
%\right.
%\\
%\left.
-6\,\mu_0\!\left(\mu_0^2-\frac{3}{5}\right)\right] \zeta_+\,\zeta_-
\Abst{,}
\end{multline}
%-----------
for the third order in $\bd$. The total motion induced y-parameter is
then given by the sum $y=\yc+y_{\rm r, oct}$, where $\yc$ is given by
\eqref{eq:yc}. As before in the limit $\thr\rightarrow 0$ the
y-parameter $y_{\rm r, oct}$ vanishes. It has a dipole and octopole
angular dependence and reaches a maximal value of $y_{\rm r, oct}\sim
9.6\cdot 10^{-11}$ in the $\mu_0, \thr$-plane. Therefore one may
completely neglect the contribution of the second beam moment to the
y-parameter. If we consider the third beam moment we find
%-----------
\bmulti
\label{eq:D3}
\left<\delta^3 \right>_{\rm r}=6\,\yd\,\bd \cdot
\left[-\frac{2}{5}\,\mu_0+\mu_0\!\left(\mu_0^2-\frac{3}{5}\right)\right] \zeta_+\,\zeta_-^2
\Abst{.}
\end{multline}
%-----------
It reaches a maximal value of $\left<\delta^3 \right>_{\rm r}/6\sim
1.8\cdot 10^{-11}$. Due the strong dependence on $\thr$ we may
therefore also neglect any contribution of the third beam moment.

%-----------
\section{Spectral distortions induced in differential measurements of
  the CMB sky}
\label{sec:strat}
\begin{figure}
\centering
%\resizebox{\hsize}{!}{\includegraphics[width=8.5cm]{./eps/yplot_1.eps}}
\includegraphics[width=\plotwd]
%\plotone
{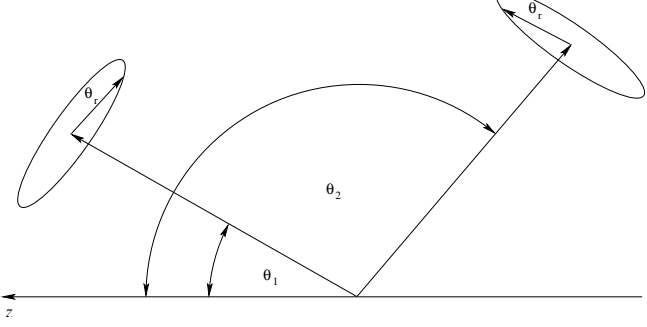}
\caption{Illustration of a differential observing strategy: The
  maximum of the dipole lies in the direction of the $z$-axis. Both
  observed patches have the same radius $\thr$ and the observed
  intensity difference is $\Delta I=I_2-I_1$.}
\label{fig:Strat_B}
\end{figure}

%-----------
In Sect. \ref{app:dipole} we discussed the spectral distortion arising
due to the CMB dipole {\it inside} a single circular beam in
comparison to a {\it reference} blackbody with temperature $T_0$.
%--------------------------------------
In this chapter we address the spectral distortion arising in {\it
differential} measurements of the CMB fluctuations, where two beam
intensities $I_1$ and $I_2$ are directly compared with each other and
the intensity difference $\Delta I=I_2-I_1$ is measured. Since in
Sect. \ref{sec:yWMAP} we have shown that y-distortion arising due to
higher multipoles have corresponding y-parameters $y\leq 10^{-9}$,
here we are only taking distortions arising due to the CMB dipole into
account.
%---------------------------------------------------
If we assume that the both beams are circular and have the same radius
$\thr$ (see Fig. \ref{fig:Strat_B}), we may define the beam RJ
temperatures as $\bar{T}_{\rm r, 1}$ and $\bar{T}_{\rm r, 2}$, where
$\bar{T}_{\rm r,\it i}$ of each beam is given by equation
\eqref{eq:T_m_d}. 
%---------------
%---------------
Now, defining the reference blackbody $I_\ast=\Ipl(\Tast)$ with
temperature $\Tref=\Tast$ using equations \eqref{eq:DT_DI} and
\eqref{eq:Sup_plancks} we find for the inferred temperature difference
at frequency $\x_{\ast}=h\nu/\kB \Tref$
%-----------
\beal
\label{eq:DI_B}
\frac{\Delta T(\x_{\ast})}{\Tast}
&=\Deltars+\ydifrs\cdot g_y(\x_{\ast})
\Abst{,}
\end{align}
%------------
where we have introduced the abbreviations
%-----------
\bsub
\label{eq:Deltas_diff}
\beal
\label{eq:Deltas_diffa}
\Deltars&=\frac{\bar{T}_{\rm r,2}-\bar{T}_{\rm r,1}}{\Tast}
\\[1mm]
\label{eq:Deltas_diffb}
\ydifrs&=\ybr+\yref
\\[1mm]
\label{eq:Deltas_diffc}
\yref&=\frac{\bar{T}_{\rm r,2}-\bar{T}_{\rm r,1}}{\bar{T}_{\rm r,1}+\bar{T}_{\rm r,2}}
\cdot\frac{\bar{T}_{\rm r,1}+\bar{T}_{\rm r,2}-2\,\Tast}{\Tast}
\Abst{.}
\end{align}
\esub
%------------
Using equation \eqref{eq:yc}, the difference of the beam y-parameters
$\ybr=\yr(\mu_2)-\yr(\mu_1)$ can be written as
%-----------
\beal
\label{eq:y_b_r}
\ybr&=-3\,\yd\,\Delta\mu\,[2\,\mu_1+\Delta\mu]\,\zeta_+\,\zeta_-
\Abst{,}
\end{align}
%------------
where $\mu_i=\cos \theta_i$, $\Delta\mu=\mu_2-\mu_1$ and $\theta_i$ is
the angle between the dipole axis and the beam $i$. Here it is
important to note, that in the case $\Delta \mu=2$, i.e. when we are
comparing the maximum and minimum of the CMB dipole $\ybr\equiv 0$,
since both beams have the same shape and size. In this case we are
directly dealing with the difference of two blackbodies with
temperatures $\bar{T}_{\rm r,\it i}$ and we may apply the results
obtained earlier in Sect. \ref{sec:Sup2P}.
%Of course in this situation distortions due to higher multipoles will
%introduce residual distortions but on a completely negligible level.

Now one can write for the relative temperature difference of the two
beams at two frequency $\x_{\ast,1}$ and $\x_{\ast,2}$, with
$\x_{\ast,2}>\x_{\ast,1}$ and $\x_{\ast,i}=h\nu_i/\kB \Tast$ as
%-----------
\beal
\label{eq:DDI_B}
\Delta\Delta
&=\frac{\Delta T(\x_{\ast, 2})-\Delta T(\x_{\ast, 1})}{\Tast}
\nonumber\\[1mm]
&
=\ydifrs\cdot[g_y(\x_{\ast, 2})-g_y(\x_{\ast, 1})]
\Abst{.}
\end{align}
%------------
If we assume that $\x_{\ast, 1}\ll 1$, i.e. for a measurement in the
RJ region of the CMB spectrum, then we can approximate
$g_y(\x_{\ast,1})\approx \x_{\ast,1}^2/6$. For estimates we will assume
that $g_y(\x_{\ast,1})=0$. 

%-------------------------------------------------
On the CMB sky the most natural choices for the reference temperature
is the full sky mean temperature $T_0$ and the temperature of the
maximum or the minimum of the CMB dipole anisotropy, whereas the last
two are equivalent. Below we now discuss the dependence of the
inferred y-parameter on the angle between the beams for these two
choices of the reference temperature. In the limit $\thr\rightarrow 0$
we can apply the results obtained in Sect. \ref{app:Dif2P} for the
difference of two blackbodies.

%---------------
\begin{figure}
%\centering
%\resizebox{\hsize}{!}{\includegraphics[width=8.5cm]{./eps/yplot_1.eps}}
%\includegraphics[width=\plotwd]
\includegraphics[width=8.5cm]
%\plotone
%{./eps/y0_strat.eps}
%{./eps/y_strat_A.ps}
{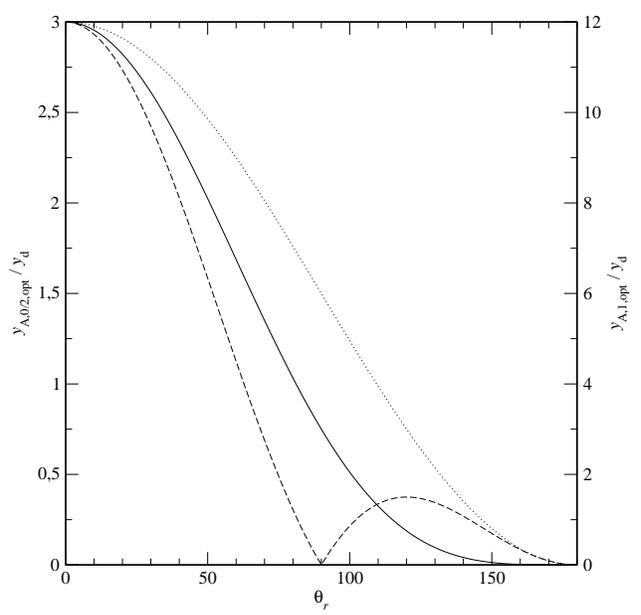}
\caption{Dependence of the y-parameter in units of $\yd=2.6\cdot
10^{-7}$ on the beam radius $\thr$ in degree for differential
measurements: $y_{\mathcal{A},0,\rm opt}$ as given by equation
\eqref{eq:y_B0opt} (dashed/left ordinate), $y_{\mathcal{A},1,\rm opt}$
as given by equation \eqref{eq:y_Bopt} (solid/right ordinate) and
$y_{\mathcal{A},2,\rm opt}$ as given by equation \eqref{eq:y_Bopt_2}
(dotted/left ordinate) .}
\label{fig:zero_differential}
\end{figure}
%------------------------
\subsubsection*{Case $\Tast=T_0$}
In this case, making use of equation \eqref{eq:T_m_d} and
\eqref{eq:Deltas_diffc} one may find
%-----------
\beal
\label{eq:y_ref_0}
y_{\rm ref, 0} &=3\,\yd\,\Delta\mu\,[2\,\mu_1+\Delta\mu]\,\zeta_+^2 +\mathcal{O}(\bd^3)
\Abst{.}
\end{align}
%------------
Using the beam y-parameter $\ybr$ as given in equation
\eqref{eq:y_b_r} we find for the corresponding total inferred
y-parameter
%-----------
\beal
\label{eq:y_B1}
y_{\rm \mathcal{A},0}=3\,{\yd}\left|\mu_{\rm r}\,\zeta_+\,\Delta \mu\cdot[2\,\mu_1+\Delta\mu]\right|+\mathcal{O}(\bd^3)
\Abst{.}
\end{align}
%------------
The largest values of the y-parameter in the $\mu_1, \Delta\mu$-plane
are obtained for the combinations $(\mu_1, \Delta\mu)=(0,1)$,
$(0,-1)$, $(1,-1)$, $(-1,1)$. In these optimal cases it follows
%-----------
\beal
\label{eq:y_B0opt}
y_{\mathcal{A},0,\rm opt}=3\,{\yd}\,\zeta_+\left|\mu_{\rm r}\right|+\mathcal{O}(\bd^3)
\Abst{.}
\end{align}
%-----------
The dependence of $y_{\mathcal{A},0,\rm opt}$ on the radius of the
beam is shown in Fig. \ref{fig:zero_differential}. The distortion does
not vanish for high angular resolution ($\theta_{\rm r}\rightarrow 0$)
and the maximum is 3 times bigger than ${\yd}$. The corresponding
absolute temperature difference in some of the {\sc Planck} frequency
channels are given in Table \ref{tab:one}. For $\thr=\pi/2$ the
y-distortion vanishes.

Let us note here, that in the case $\Delta\mu=2$, i.e. when we are
directly comparing the maximum and the minimum of the CMB dipole, the
spectral distortion vanishes. This can be understood as follows: As
was argued in Sect. \ref{sec:SupP}, if the chosen reference
temperature is equal to $T_{y=0}$ as given by equation
\eqref{eq:Tref_y_is_0} the total y-parameter is zero. For $\mu_1=-1$
and $\mu_2=1$ the beam moments $\left<\delta_i^2\right>_{\rm r,\it i}$
are both equal and therefore do not contribute to the total inferred
y-parameter \eqref{eq:Deltas_diffb}. In this case we obtain
$T_{y=0}=\TRJ=(\bar{T}_{\rm r,1}+\bar{T}_{\rm
r,2})/2=T_0+\mathcal{O}(\bd^2)$. Therefore, consistent in second order
of $\bd$ the total inferred y-parameter vanishes. This conclusion can
also be directly drawn from equation \eqref{eq:y_inf_gen}.

\subsubsection*{Case $\Tast=\Tbar_{\rm r, 1}$}
In this case, again making use of equation \eqref{eq:T_m_d} and
\eqref{eq:Deltas_diffc} one may find
%-----------
\beal
\label{eq:y_ref_1}
y_{\rm ref, 1} &=3\,\yd\,\Delta\mu^2\,\zeta_+^2+\mathcal{O}(\bd^3)
\Abst{.}
\end{align}
%------------
Now using equation \eqref{eq:y_b_r} we may write the total inferred
y-parameter as
%-----------
\beal
\label{eq:y_B2}
y_{\rm \mathcal{A},1}=3\,{\yd}\zeta_+\,\left|\Delta\mu\cdot[\mu_{\rm r}\,\mu_2-\mu_1]\right|+\mathcal{O}(\bd^3)
\Abst{.}
\end{align}
%-----------
The largest values of the y-parameter in the $\mu_1, \Delta\mu$-plane
are obtained for the combinations $(\mu_1, \Delta\mu)=(1,2)$ and
$(-1,-2)$. In these optimal cases the y-parameter is
%-----------
\beal
\label{eq:y_Bopt}
y_{\mathcal{A},1,\rm opt}=12\,{\yd}\,\zeta^2_+ +\mathcal{O}(\bd^3)
\Abst{.}
\end{align}
%-----------
For high resolution ($\zeta_+\rightarrow 1$) we can obtain
$y_{\mathcal{A},1,\rm opt}$ directly as the difference of two Planck
spectra with temperature corresponding to the maximum $T_2$ and the
minimum $T_1$ of the dipole using \eqref{eq:DD2} with $\Tref=T_1$. In
this case we directly get
$\frac{1}{2}\,[(T_2-T_1)/T_1]^2=2\,\bd^2=12\,\yd$. 
%----------------------
As will be discussed later this way of comparing the maximum and
minimum of the CMB dipole should open a way to cross calibrate the
frequency channels in future experiments with full sky coverage to
very high precision (Sect.~\ref{sec:Cali}).
 
The dependence of $y_{\mathcal{A},1,\rm opt}$ on the beam radius is
shown in Fig. \ref{fig:zero}. As in the previous case, the
y-distortion does not vanish for high resolution ($\theta_{\rm
r}\rightarrow 0$) and in the maximum it is even 12 times bigger than
${\yd}$, corresponding to $\yo=3.1\cdot 10^{-6}$. This is only $\sim
5$ times below the current upper limit on the mean y-parameter given
by {\sc Cobe/Firas}. The corresponding absolute temperature difference
in some of the {\sc Planck} frequency channels are given in Table
\ref{tab:one}. 

%-------------------------------------------------
Another case may be interesting for CMB missions with partial sky
coverage, but which have access to the maximum or the minimum of the
dipole ($\mu_1=\pm 1$) and a region in the directions perpendicular to
the dipole axis ($\mu_2=0$): Using \eqref{eq:y_B2} with $(\mu_1,
\Delta\mu)=(\pm 1,0)$ one obtains
%-----------
\beal
\label{eq:y_Bopt_2}
y_{\mathcal{A},2,\rm opt}=3\,{\yd}\,\zeta_+ +\mathcal{O}(\bd^3)
\Abst{.}
\end{align}
%-----------
In this case the amplitude is comparable to $y_{\rm \mathcal{A},0}$,
but the dependence on the beam radius $\thr$ is much weaker (see
Fig. \ref{fig:zero}).

\section{Cross Calibration of frequency channels}
\label{sec:Cali}
%-----------
The cross calibration of the different frequency channels of CMB
experiments is crucial for the detection of any frequency dependent
signal. As has been mentioned earlier usually the dipole and its
annual modulation due to the motion of the earth around the sun is
used for calibration issues. The amplitude of the dipole is known with
a precision of $\sim 0.3\%$ on a level of $\Delta T\sim\,$mK. But the
sensitivities of future experiments will be significantly higher than
previous missions and request a possibility for cross calibration down
to the level of tens of nK.

In this Sect. we now discuss two alternative ways to cross calibrate
the frequency channels of future CMB experiments. Most importantly we
show that making use of the superposition of blackbodies and the
spectral distortions induced by the CMB dipole should open a way to
cross calibrate down to the level necessary to detect signals from the
dark ages as proposed by \citet{Kaustuv2003}.

\subsection{Calibration using clusters of galaxies}
\label{sec:CaliClust}
It is obvious, that brightest relaxed clusters of galaxies on the sky
open additional way to cross calibrate the different frequency
channels of future CMB experiments like {\sc Planck} and ground based
experiments like {\sc Act}.
%-------------------------------------------------------------------
The CMB signal of the majority of rich clusters has very distinct
y-type spectrum due to the thermal SZE. Multifrequency measurements
allow to determine their y-parameters and thereby open a way to use
them as sources for cross calibration. This cross calibration will
then enable us to detect the effects of relativistic correction for
very hot clusters and thereby to measure the temperatures of clusters
independent of X-ray observations.

%--------------------------------------------------------------
Clusters such as Coma, where existing X-ray data provide excellent
measurements of the electron densities and temperatures inside the
cluster and where we know the contamination of the CMB brightness by
radiosources and the dust in galaxies very well, even may be used for
absolute calibration of CMB experiments. For the Coma cluster today
X-ray data allows us to predict the CMB surface brightness with
precision of the order of a few percent.
%--------------------------------------------------------------
%Using existing samples of clusters \citep{Reese2002} will
%statistically decrease the uncertainties for example due to
%asphericities, projection effects and the kinetic SZE and therefore
%further improve the properties of clusters as a source for absolute
%and cross calibration.

Due to the redshift independence of the SZ signal, distant clusters
have similar surface brightness but much smaller angular
diameters. Therefore they are good sources for calibration issues for
experiments covering only limited parts of the sky and having high
angular resolution like {\sc Act} (for example {\sc Act} is planing to
investigate 200 square degree of the sky \citep{Kosowski2003}).

%-----------
One disadvantage of clusters for cross calibration purposes is that
they are too bright: Typically clusters have y-parameter of the order
of $\sim 10^{-5}-10^{-4}$ and therefore are only a few times weaker
than the temperature signal of the CMB dipole ($\sim 3\,$mK). In
comparison to the level of calibration and cross calibration achieved
directly using the dipole this is not much of an improvement. 

In addition planets or galactic and extragalactic radiosources might
serve for cross calibration purposes, if we know their spectral
features with very high precision. Quasars and Active Galactic Nuclei
are not good candidates for this issue, since they are highly variable
in the spectral band of interest. 

\subsection{Cross calibration using the superposition of blackbodies} 
\label{sec:CaliSup}
%-------------------------------------------------------
All the sources mentioned in Sect. \ref{sec:CaliClust} do not allow
cross calibration down to the level of the sensitivity of future CMB
experiments and do not enable us to detect signals from the dark ages
as predicted by \citet{Kaustuv2003}. Therefore here we want to discuss
a method to cross calibrate the frequency channel using the spectral
distortions induced by the CMB dipole (see
Sect. \ref{app:dipole}). These spectral distortions can be predicted
with very high accuracy on a level of $\sim \mu$K. This is $\sim 1000$
times lower than the dipole signal and should therefore allow cross
calibration to a very high precision.

%-------------------------
Every CMB experiment is measuring the fluctuations of the CMB
intensity on the sky. Due to the nature of the physical processes
producing these fluctuations in each point they are related to the
fluctuations of the radiation temperature.
%-------------------------------------------------
In a map making procedure the measured fluctuations in the intensity
are translated into the fluctuations in the radiation
temperature. Usually the aim of any map making procedure is to reduce
statistic and systematic error in order to obtain a clean signal from
these CMB temperature fluctuations.
%--------------------------------------------------
As has been discussed in Sects. \ref{app:Theorie} the superposition of
blackbodies with slightly different temperature in second order
induces y-distortions in the {\it inferred} temperature
differences. The amplitude of these distortions depends on the
temperature difference and the chosen reference temperature.
%---------------
In Sects. \ref{app:dipole} and \ref{sec:strat} it has been shown that
due to the two basic observing strategies, {\it absolute} and {\it
differential} measurements, the dipole can induce y-distortions with
y-parameters up to $y\sim 10^{-6}$. It was also shown in
Sect. \ref{sec:yWMAP} that contributions from higher multipoles are
much smaller ($y\sim 10^{-11}-10^{-9}$).

The y-distortions induced by the dipole can contaminate the maps
produced in the map making procedure on a level higher than the
sensitivity of the experiment. Therefore it is important to chose the
map making procedure such that spectral distortions are minimized. The
discussion in Sects. \ref{sec:Sup2P} has shown, that for this purpose
in {\it absolute} measurements of the CMB sky it is the best to choose
the temperature of the reference blackbody close to the beam RJ
temperature. This implies that for CMB experiment the best choice for
the temperature of the internal calibrator is the full sky mean
temperature $T_0$.
%--------------------------------------------------
In order to minimize the spectral distortions arising due to the
superposition of blackbodies in {\it differential} measurements the
best choice for the reference temperature used to relate the intensity
difference maps to the temperature difference maps is the RJ
temperature of the combined temperature distribution of both
beams. This optimal reference temperature will be time dependent due
to the various scanning strategies and orientations of the spinning
axis relative to the dipole during observations.

%-------------------------
The purpose of this Sect. is not to discuss details about map making
procedures but to show, that there are ways to manipulate these CMB
maps in order to make the spectral distortions arising due to the
superposition of blackbodies become useful for cross calibration
purposes.
%-----------------------------------------------------
As the discussion in Sect. \ref{sec:Sup2P} has shown, for this issue
it is better to move $\Tref$ as far as possible away from $T_0$. The
optimal choice of $\Tref$ depends on the sensitivity of the experiment
and on the observing strategy: For {\it absolute} measurements $\Tref$
should be as close as possible to $T_0$ in order to minimize the
induced y-distortions but on the other hand it should be chosen such
that the induced y-distortion due to the CMB dipole is still
measurable within the sensitivity of the experiment. But in principle
there is no strong constrain, since the induced y-distortions can
again eliminated afterwards.
%----------------------------------------------
For {\it differential} measurements it is possible to chose the
optimal reference temperature for calibration issues independent of
the best map making reference temperature, but here it is important to
compare regions on the sky with maximal temperature difference,
i.e. with maximal angular separation in the observed field.
%----------------------------------------------
Below we now separately discuss CMB experiments using {\it
differential} measurements with full and partial sky coverage in more
detail.

\subsection*{Full sky CMB surveys}
For full sky missions like {\sc Planck} the regions with maximal
temperature difference are located around the extrema of the CMB
dipole. Using equation \eqref{eq:y_B2} we see that they correspond to
$\Delta \mu=2$ and the resulting maximal spectral distortion is
characterized by $\yo=12\,\yd=3.1\cdot 10^{-6}$, if we set
$\Tref=\Tbar_{\rm r, 1}$ or $\Tref=\Tbar_{\rm r, 2}$, i.e. to the
maximum or minimum of the temperature on the CMB sky arising due to
the dipole.

%---------------
\begin{table}%[h]
\centering
\caption{y-distortion: $\Delta T=\Tref\cdot y$ in $\mu$K 
for $y$ as given in the left column 
in some of the {\sc Planck} spectral channels. Here $\yd=2.6\cdot
10^{-7}$, $y_{\rm opt, 0}=3\,{\yd}$ and $\yo=12\,{\yd}$.}
%---------------
\begin{tabular}{lllllllll}%l}
\\
& \multicolumn{7}{c}{Center of channels [GHz]}
%\\{[$\mu$K]} & \multicolumn{8}{c}{[GHz]}
\\
\hline\\[-8pt]
$\nu_{\rm c}$ & 30 & 44 & 70 & 100 & 143 & 217 & 353 %& 547
\\[1.2mm]
\hline\hline\\[-8pt]
$\yd$ & 0.03 & 0.07 & 0.17 & 0.34 & 0.67 & 1.39 & 2.97 %& 5.34
\\
$y_{\rm opt, 0}$ & 0.10 & 0.21 & 0.52 & 1.03 & 2.01 & 4.18 & 8.90 %& 16.02
\\
$\yo$ & 0.39 & 0.83 & 2.07 & 4.13 & 8.05 & 16.71 & 35.56 %& 64.00
\\
%\hline\\[-8pt]
%{\sc Wmap} & 23 & 33 & 41 & 61 & 93 & & & 
%\\[1.2mm]
%---------------
\label{tab:one}
\end{tabular}
\end{table}
%---------------

One attractive procedure to compare the maximum and minimum of the CMB
dipole is to take the CMB intensity maps of each spectral channel and
to calculate the difference between each map and the map obtained by
remapping the value of the intensity $I(\vek{n})$ at position
$\vek{n}$ to the value $I(-\vek{n})$ at position $-\vek{n}$, i.e
rotating the initial map by $180$ degrees around any axis, which is
perpendicular to the CMB dipole axis and is crossing the origin or
equivalently setting $\mu_2=-\mu_1$ in equation \eqref{eq:DI_B}. 
%In the following we will refer to this transformed map as the {\it
%rotated} map.

Afterwards these artificial intensity maps for each spectral channel
are converted into temperature maps using equation \eqref{eq:DT_DI}
and setting the reference temperature at each point to
$\Tast(\vek{n})=\bar{T}_{\rm r,1}(\vek{n})$ of the intrinsic
map. Neglecting the contributions from intrinsic multipoles with
$l\geq 2$, the difference map in some frequency channel
$\x_\ast=h\nu/\kB \bar{T}_{\rm r,1}$ will be given by
%-----------
\bmulti
\label{eq:DI_trans}
\frac{\Delta T(\x_{\ast},\vec{n})}{\bar{T}_{\rm r,1}(\vec{n})}
=-2\,\bd\,\zeta_+\,\mu+2\,\bd^2\,\zeta_+^2\,\mu^2
\\
+12\,\yd\,\zeta^2_+\,\mu^2\cdot g_y(\x_{\ast})
\Abst{,}
\end{multline}
%------------
where we used the definition \eqref{eq:DT_T_y} for $g_y(\x)$.  This shows, that
the artificial maps will contain a frequency independent temperature
dipole component with twice the initial dipole amplitude and a
monopole and quadrupole y-distortion resulting from the second term in
\eqref{eq:DI_trans}. The maxima of the spectral distortion will
coincide with the extrema of the CMB dipole.

Now, using the artificial map of the lowest frequency channel as a
reference and taking the differences between this reference map and
the artificial maps in higher spectral channels we can eliminate the
frequency independent term corresponding to twice the dipole in
equation \eqref{eq:DI_trans}. This then opens a way to cross calibrate
all the channels to very high precision, since the quadrupole
component in these artificial maps will be larger in higher spectral
channels, following the behavior of $g_y$ with $\x$.

The signal will include both statistical and systematic errors for the
average temperature of the sky and the dipole amplitude $\bd$. All
uncertainties are mainly influencing the frequency independent dipole
term but for the frequency dependent monopole and quadrupole they will
become important only in next order:
%-----------
\beal
\label{eq:error}
\bar{T}_1\cdot\bd^2
&=\bar{T}_{1,0}\,\beta_{\rm d, 0}^2\,(1\pm\Delta_T)\cdot(1\pm\Delta_\beta)^2
\nonumber\\
&=\bar{T}_{1,0}\,\beta_{\rm d, 0}^2\,[1\pm 2\,\Delta_\beta\pm\Delta_T]
+\mathcal{O}(\Delta_\beta\cdot\Delta_T)
\Abst{,}
\end{align}
%------------
where $\bar{T}_{1,0}$ is the correct value of the temperature at a
given point on the sky, $\beta_{\rm d, 0}$ is the correct dipole
amplitude and $\Delta_T$ and $\Delta_\beta$ are their corresponding
relative uncertainties. This implies that all the corrections to the
frequency dependent terms are at least 1000 times smaller than the
signal we are discussing here. 
%----------------------------------
On this level correction due to higher order moments might become
important, but their contributions will be less than 1\%. This means
that y-distortion on a level of $10\,\mu$K can be predicted with $\sim
100\,$nK precision.

Let us note here, that applying the procedure as described above we are
adding signals which have statistically independent noise. Therefore
the statistical noise of the artificial maps should be a factor
of $2/\sqrt{2}=\sqrt{2}$ stronger than the initial maps. On the other
hand we are increasing the amplitude of the y-distortion quadrupole by
a factor of $12$, resulting in strong gain.

%-------------------------------------------------------
The maxima of the quadrupole component in the artificial maps will be
rather broad and correspond to thousands of square degrees on the sky,
allowing to average the signal and thereby increasing the sensitivity.
For circular beam average formula \eqref{eq:DI_trans} will be
applicable. The statistical sensitivity of the {\sc Planck} experiment
will permit to find these spectral distortions in the artificial maps
down to the level which is necessary to detect the effects of
reionization as discussed by \citet{Kaustuv2003}. Applying the above
procedure to the $33\,$GHz and $93\,$GHz frequency channels of {\sc
Wmap} will lead to a maximal temperature difference of $\Delta
T=3.6\,\mu$K, whereas the maximal difference will be $\Delta T=
35.6\,\mu$K for the $353\,$GHz channel of {\sc Planck} (see Table
\ref{tab:one}). Nevertheless, even in the case of {\sc Wmap} the very
high precision of the experiment might permit the detection of this
quadrupole and thereby open the way to cross calibrate its different
spectral channels.

For {\sc Cobe/Firas} the proposed method permits to check the
precision of the internal calibration. The internal calibrator was
measured and tested with some finite precision and there is the
possibility that the internal calibration may be better than the
precision at which it was tested. The proposed method has distinct
spectral and angular properties, making it possible to improve the
cross calibration of the spectral channels and possibly this will open
a way to further improve the great results this experiment already
gave us.

\subsection*{CMB surveys with partial sky coverage}
%The sensitivities of {\sc Act} or the next generation multichannel
%{\sc Cbi} experiment will in principle allow to detect signatures from
%the dark ages as discussed by \citet{Kaustuv2003}. However these
%experiments will only cover restricted regions of the sky (for example
%{\sc Act} is planing to investigate 200 square degree
%\citep{Kosowski2003}).

%Lyman Page (priv. com.) mentioned that the main problem for the
%observation of frequency dependent signals from the dark ages as
%discussed by \citet{Kaustuv2003} will be the necessity to cross
%calibrate different frequency channels with accuracy better than the
%signal. 

%In order to use the method of cross calibration as proposed above, it
%is the best to separate the area under investigation into two equal
%well separated regions on the sky with largest possible $\Delta
%\mu$. The dependence of the signal on the seperation angle has been
%discussed in detail in Sect. \ref{sec:strat}.
%Figure XXX shows the dependence of the signal for two choices of the
%reference field on the angular separation of the second field.

Surveys with partial sky coverage may not simultaneously have access
to regions around the maximum and the minimum of the dipole. Therefore
here one should choose two regions in the accessible field of view
such that the mean temperature difference between these regions is as
large a possible. The dependence of the signal on the separation angle
has been discussed in detail in Sect. \ref{sec:strat}.

Before cross calibration all the bright sources detected such as
bright clusters of galaxies and point sources inside the chosen areas
should be extracted. Typically the expected amount of SZ sources will
be of the order of a few tens per square degrees with corresponding
angular extensions less than $1'$ \citep[see][]{CarlstromXXX,
Jose2003}. Afterwards the difference of the signals in the two patches
can be taken and compared to the signals in the other spectral
channels.

%--------------------------------------------------------
For example choosing one area centered on the minimum or maximum of
the CMB dipole as a reference ($\Tast=\bar{T}_{\rm r,1}$) and choosing
the second area in the ring perpendicular to the dipole axis the
maximal y-parameter is $y=3\,\yd\,\zeta_+=7.7\cdot 10^{-7}\,\zeta_+$
(cf. equation \eqref{eq:y_Bopt_2}). If we instead set $\Tast=T_0$ we
obtain the same maximal y-parameter but a stronger dependence on the
beam radius or size of the regions we average (cf. equation
\eqref{eq:y_B0opt}). This small example shows that for any experiment
with partial sky coverage a separate analysis of the optimal choice of
the reference temperature and the regions on the sky has to be done.

%-----------
%\section{Small scale density perturbations dissipated before and during recombination}
\section{Other sources of spectral distortions}
%-----------
\label{sec:conseq}
Injection of energy into the CMB prior to recombination leads to
spectral distortions of the background radiation. Given the \WMAP best
fit parameters, before redshift $z_{\rm th}\sim 2\cdot 10^6$ all
distortions are efficiently wiped out, whereas energy injection in the
redshift range $z_{\rm th}> z > z_{\mu}$, with $z_{\mu}\sim 1.4\cdot
10^5$, leads to a $\mu$-distortion and injection in the range $z_{\mu}
> z > z_{\rm rec}$ to a y-distortion, with $z_{\rm rec}\sim 1090$
\citep{Suny70b,Ill75, Burigana91,Hu93}.
%Another y-distortion arises from the reionization of the intergalactic medium.

As was shown by \cite{Silk68} photon diffusion and thermal viscosity
lead to the dissipation of small scale density perturbations before
recombination. This damping of acoustic waves will contribute to a
$\mu$- and y-distortion because it was leading to (i) energy release
\citep{Suny70b,Daly91, Hu94} and (ii) mixing of photons from regions
having different temperatures \citep{Zeld72}.

Now, the \WMAP data implies that the initial spectrum of perturbations
is close to a scale-invariant Harrison-Zeldovich spectrum with
spectral index $n=1.03\pm 0.04$ \citep{WMAP_params}. Different
estimates for the spectral distortions arising from the dissipation of
acoustic waves in the early universe give a chemical potential of the
order of $\mu_{\rm dis}\sim 2\cdot 10^{-7}$ and y-distortions with
$y_{\rm dis}\sim 10^{-7}$ \citep{Daly91,Hu94}.

Another contribution to the y-distortion arises from the epoch of
reionization \citep{Zeld69,Hu94b} and from the sum of the SZE of
clusters of galaxies \citep{Mark91}. The \WMAP results for the TE
power spectrum point towards an early reionization of the universe
with corresponding optical depth $\tau_{\rm re}=0.17\pm 0.04$
\citep{Kogut03, Spergel03}. Here two main effects are important: (i)
The photoionized gas typically has temperatures $T$ of the order of
$\sim 10^4\,$K. Therefore the diffuse gas after reionization will
produce a y-distortion in the CMB spectrum with
%-----------
%-----------
\beal
%\label{eq:yre}
y_{\rm re}\sim \int \frac{\kB T}{\me\,c^2}\,\id\tau \sim 3\cdot 10^{-7}\;
\frac{\tau_{\rm re}}{0.17}\;
\frac{T}{10^4\,\text{K}}\nonumber
\Abst{.}
\end{align}
%-----------
%-----------
(ii) The motion of the matter induces a y-distortion on the whole sky due to
the second order Doppler effect, with corresponding y-parameter \citep{Hu94b},
%-----------
$y_v=\frac{1}{3}\int \sigT\,\Ne\,c\left<v^2\right>\id t\nonumber$,
%-----------
where $\left<v^2\right>$ is the velocity dispersion over the whole sky
in units of the speed of light. For a CDM model it is of the order of
%-----------
\beqa
\label{eq:yvv}
\qquad \quad y_v\sim 2\cdot 10^{-8}\left(\left[\frac{z_{\rm re}}{10}\right]^{1/2}-1\right)
\frac{\left<v^2\right>}{10^{-5}}\,\frac{\Obhh}{0.0224}\nonumber
\Abst{.}
\eeqa
%-----------
Since the reionization optical depth is very small, this is one order
of magnitude smaller than $y_{\rm re}$. If secondary ionization was
patchy this will in addition lead to angular fluctuations
\citep{Santos2003}.

All the effects mentioned in this section lead to y-distortions with
corresponding y-parameter in the range $y\sim 10^{-8}-10^{-7}$. As has
been shown in Sects. \ref{app:dipole} and \ref{sec:strat} the
y-distortions arising due to the CMB dipole are of the same order. For
example the CMB dipole induces a full sky y-distortion with
${\yd}=2.6\cdot10^{-7}$. Therefore, to measure any of the effects
discussed above it is necessary to take spectral distortions
associated with the CMB dipole into account. But since the angular
distribution and the amplitude of these distortions can be accurately
predicted they can be easily extracted. As has been shown in
Sect. \ref{sec:yWMAP} spectral distortions arising from higher
multipoles typically have $y\sim 10^{-11}-10^{-9}$ and therefore do
not play an important role in this context. But even for these
distortions the locations and amplitudes can be accurately predicted
with measured CMB maps and therefore offer the possibility to
eliminate these distortions from the maps.

\section{Conclusion}
\label{sec:conc}
We have discussed in detail the spectral distortions arising due to
the superposition of blackbodies with different temperatures in the
limit of small temperature fluctuations. The superposition leads to
y-distortion with y-parameter $y_{\rm S}=\left<\delta^2\right>/2$,
where $\left<\delta^2\right>$ denotes the second moment of the
temperature distribution function, if the difference in temperatures
in less than a few percent of the mean RJ temperature. We have shown
that in this limit even comparing to pure blackbodies leads to a
y-distortion.

%-----------
The results of this derivation where then applied to measurements of
the CMB temperature anisotropies with finite angular resolution and in
particular to the CMB dipole and its associated spectral distortions,
but in principle the method developed here can be applied whenever one
is dealing with the superposition of blackbodies with similar
temperatures. We have shown, that taking the difference of the CMB
intensities in the direction of the maximum and the minimum of the CMB
dipole leads to a y-distortion with y-parameter $\yo=3.1\cdot
10^{-6}$. This value is 12 times higher than the y-type monopole,
$\yd=2.6\cdot 10^{-7}$. Since the amplitude of this distortion can be
calculated with the same precision as the CMB dipole, i.e. $0.3\%$
today \citep{Fixsen02}, it opens a way to cross calibrate the
different frequency channels of CMB experiment down to the level of a
few tens of nK ( for more details see Sect. \ref{sec:Cali}).

%-------------------------------
We discussed another possibility to check the zero levels of different
frequency channels by observing the difference of the brightness in
the direction of the dipole maximum and in the direction perpendicular
to the dipole axis: The dipole induced spectral distortion in this
case is 4 times weaker than for the difference of the maximum and
minimum and corresponds to $y\sim 7.7\cdot 10^{-7}$. Nevertheless, it
is still 3 times stronger than the dipole induced whole sky
y-distortion with y-parameter $\yd$.

%-------------------------------
The value of $\yo$ is only 5 times lower than the upper limit on the
whole sky y-parameter obtained by the {\sc Cobe/Firas} experiment
\citet{Fixsen02} and is orders of magnitudes above the sensitivity of
{\sc Planck} and {\sc Cmbpol} in each of their spectral
channels. Therefore this signal might become useful for both cross
calibration and even absolute calibration of different frequency
channels of these experiments to very high precision, in order to
permit detection of the small signals of reionization as for example
discussed by \citet{Kaustuv2003}, and to study frequency dependent
foregrounds with much higher sensitivity. We should emphasize, that on
this level we are only dealing with the distortions arising from the
CMB dipole as a result of the comparison of Planck spectra with
different temperatures close to the maximum and minimum of the CMB
dipole, i.e. due to the superposition of blackbodies. The distortions
are introduced due to the processing of the data and become most
important in the high frequency channels.

Let us stress that the distortions discussed in this work have both
well known spectral and angular dependence and that they are connected
with the much stronger signal of the CMB dipole. The amplitude of the
dipole is known to a relative precision $\sim 10^{-3}$ and therefore
the induced spectral distortions can be calculated to the same
accuracy. Maybe the great properties of the calibration source
discussed above might even be detectable in the high frequency
channels ($x\gtrsim 10$) of the existing {\sc Cobe/Firas} data and thereby
help to further improve its calibration.

The development of the CMB experimental technology is repeating with
20 years delay the progress that was made in the astronomical
observations with X-ray grazing incidence mirrors and CCD
detectors. Due to the big efforts of many spacecraft teams, the X-ray
background today is resolved to more than 85 \%
\citep{Rosati2002}. During the next decade, CMB observers will be able
to pick up all rich clusters of galaxies and all bright y-distorted
features connected with supernovae in the early universe, groups of
galaxies and even rising down to patchy reionization in the CMB maps.
Deep source counts may permit the separation of their contribution to
the y-parameter and in principle might bring us close to the level of
the y-distortions mentioned in Sect. \ref{sec:conseq}. As has been
argued, on this stage the contributions to the y-parameter arising
from the superposition of Planck spectra with different temperatures
corresponding to the observed CMB temperature fluctuations will be the
easiest to separate. Again, the main contribution will arise from the
CMB dipole. On the full sky the corresponding y-parameter will be
$y_{\rm d}=2.6\cdot10^{-7}$. But even the distortions arising from the
primordial temperature fluctuations with multipoles $l\geq 2$ can lead
to distortions of the order of $y\sim 10^{-9}$ in significant parts of
the sky. In the future this type of spectral distortion can be taken
into account making it possible to enter the era of high precision CMB
spectral measurements.

%We want to note, that after this paper was close to be finished M.
%Kamionkowski mentioned an earlier publication in which some results of
%this work were already discussed \citep{Kam03}.

\acknowledgements{Some of the results in Sect.\ref{sec:yWMAP} have
been obtained using the {\sc Healpix} distribution
\citep{Gorski99}. J.C. would like to thank
C. Hern{\'a}ndez-Monteagudo, S.Yu. Sazonov,
J.A. {Rubi{\~n}o-Mart{\'\i}n}, G. Huetsi and K. Basu for very valuable
discussions and suggestions. 
%---------------------------------------------
R.S. wants to acknowledge support from the Gordon Moore distinguished
scholar fellowship which allowed working on this paper during
Dec. 2003 and Jan. 2004 at the California Institute of
Technology. R.S. also is grateful to Marc Kamionkowski for hospitality
at the Caltech, useful discussions and the information about the paper
of \citet{Kam03}. R.S. wants to acknowledge stimulating conversations
with Lyman Page, Dale Fixsen, John Mather, Charles Lawrence and Tony
Readhead especially about future experiment.}

%\begin{appendix}
%\twocolumn
%\end{appendix}

\end{document}

%% file: Befehle.tex
\newcommand{\xc}{x_{\rm c}}

\newcommand{\beq}{\begin{equation}}   %
\newcommand{\eeq}{\end{equation}}   %

\newcommand{\beqa}{\begin{eqnarray}}   %
\newcommand{\eeqa}{\end{eqnarray}}   %
\newcommand{\beal}{\begin{align}}
\newcommand{\enal}{\end{align}}
\newcommand{\bspl}{\begin{split}}
\newcommand{\espl}{\end{split}}
\newcommand{\bsub}{\begin{subequations}}
\newcommand{\esub}{\end{subequations}}
\newcommand{\bmulti}{\begin{multline}}   %
\newcommand{\beqm}{\begin{mathletters}}   %
\newcommand{\eeqm}{\end{mathletters}}   %

\newcommand{\Abst}[1]{\,#1}

\newcommand{\kB}{k_{\rm B}}
\newcommand{\me}{m_{\rm e}}
\newcommand{\Ne}{n_{\rm e}}

\newcommand{\Te}{T_{\rm e}}

\newcommand{\sigT}{\sigma_{\rm T}}

\newcommand{\vek} [1]{\mbox{\boldmath${#1}$\unboldmath}}

\newcommand{\pd}{\partial}
\newcommand{\pAb}[2]{\frac{\displaystyle\pd #1}{\displaystyle\pd #2}}
\newcommand{\PAb}[3]{\frac{\displaystyle\pd^{#3} #1}{\displaystyle\pd {#2}^{#3}}}

\newcommand{\Abl}[2]{\frac{d #1}{d #2}}

\newcommand{\Obhh}{\Omega_{\rm b}h^2}